\documentclass[final,1p,times,number]{elsarticle}

\usepackage{lineno}
\usepackage[colorlinks=true,linkcolor=black,bookmarksopen=false]{hyperref}
\usepackage{tikz}
\usetikzlibrary{arrows.meta,calc,decorations.markings,math,arrows.meta,shapes.arrows,fadings}
\usepackage{physics}
\usepackage{tikz-feynman, amsmath}
\usepackage{amsmath}
\modulolinenumbers[5]
\usepackage{comment}
\usepackage{float}
\usepackage{units}
\usepackage{amssymb, amsfonts}
\usepackage{wasysym}
\usepackage{xcolor}

\usepackage{subcaption}
\usepackage{caption}

\DeclareMathOperator{\diag}{diag}

\newcommand{\e}{\varepsilon}

\newcommand{\vex}[1]{{\bf\mathrm{#1}}}
\newcommand{\Nabla}{{\bf\nabla}}

\newcommand{\pup}[1]{{\scriptscriptstyle{({#1})}}}

\newcommand{\tauh}{\hat{\tau}}

\newcommand{\sigh}{\hat{\sigma}}
\newcommand{\sigb}{\hat{\bf{\sigma}}}

\newcommand{\T}{\mathsf{T}}

\newcommand{\parr}{\partial}

\newcommand{\bsub}{\begin{subequations}}
\newcommand{\esub}{\end{subequations}}

\newcommand{\intl}[1]{\int\limits_{#1}}

\newcommand{\vv}{\mathsf{v}}
\newcommand{\voo}{v_{11}}
\newcommand{\dvoo}{\delta v_{11}}
\newcommand{\vtt}{v_{22}}
\newcommand{\dvtt}{\delta v_{22}}
\newcommand{\vot}{v_{12}}
\newcommand{\vto}{v_{21}}
\newcommand{\sk}{\sigma_{\mathsf{dc}}}

\journal{Annals of Physics}

\begin{document}

\begin{frontmatter}

\title{How spectrum-wide quantum criticality protects surface states of topological superconductors from Anderson localization:\\
Quantum Hall plateau transitions (almost) all the way down}

\author{Jonas~F.~Karcher}
\address{{Institute for Quantum Materials and Technologies, Karlsruhe Institute of Technology, 76021 Karlsruhe, Germany}}
\address{{Institut f\"ur Theorie der Kondensierten Materie, Karlsruhe Institute of Technology, 76128 Karlsruhe, Germany}}
\author{Matthew~S.~Foster}
\address{{Department of Physics and Astronomy, Rice University, Houston, Texas 77005, USA}}
\address{{Rice Center for Quantum Materials, Rice University, Houston, Texas 77005, USA}}

\begin{abstract}
We review recent numerical studies of two-dimensional (2D) Dirac fermion theories that exhibit an unusual
mechanism of topological protection against Anderson localization. 
These describe surface-state quasiparticles of time-reversal invariant, 
three-dimensional (3D) topological superconductors (TSCs), 
subject to the effects of quenched disorder. 
Numerics reveal a surprising connection between 3D TSCs in classes AIII, CI, and DIII,
and 2D quantum Hall effects in classes A, C, and D. 
Conventional arguments derived from the non-linear $\sigma$-model picture 
imply that most TSC surface states should Anderson localize
for arbitrarily weak disorder (CI, AIII), or exhibit weak antilocalizing behavior (DIII).
The numerical studies reviewed here instead indicate \emph{spectrum-wide 
surface quantum criticality}, characterized by robust eigenstate multifractality throughout 
the surface-state energy spectrum. In other words, there is an ``energy stack'' of critical wave functions. 
For class AIII, multifractal eigenstate and conductance analysis reveals identical statistics
for states throughout the stack, consistent with the class A integer quantum-Hall plateau transition (QHPT). 
Class CI TSCs exhibit surface stacks of class C spin QHPT states. 
Critical stacking of a third kind, possibly associated to the class D thermal QHPT, 
is identified for \emph{nematic} velocity disorder of a single Majorana cone in class DIII.
The Dirac theories studied here can be represented as perturbed 2D Wess--Zumino--Novikov--Witten sigma models;
the numerical results link these to Pruisken models with the topological angle $\vartheta = \pi$. 
Beyond applications to TSCs, all three stacked Dirac theories (CI, AIII, DIII) naturally arise 
in the effective description of dirty $d$-wave quasiparticles, relevant to the high-$T_c$ cuprates.
\end{abstract}


\end{frontmatter}


\tableofcontents


\section{Introduction}

Topological phases of non-interacting fermions are classified according to the ``10-fold way'' 
\cite{SRFL2008,Kitaev2009,SRFL2010,TSCRev2}. The same scheme 
(also referred to as the Altland--Zirnbauer or Cartan classification)
applies to a seemingly unrelated problem, that of the Anderson (de)localization in the presence of quenched disorder 
\cite{AltlandZirnbauer,BernardLeClair2002,Evers2008}. In fact, topology and disorder are closely intertwined
in condensed matter physics. In both cases, one seeks to characterize not the details of a particular
band structure or disorder configuration, but the physics that robustly persists under smooth deformations
of the Hamiltonian that preserve defining symmetries. 

In each spatial dimension, five of the ten classes admit topologically nontrivial phases.
Three of the five topological classes are characterized by an integer-valued winding number
$\nu \in \mathbb{Z}$
\cite{Kitaev2009,SRFL2010}; the other two classes in each spatial dimension have $\mathbb{Z}_2$ invariants.  
In two dimensions (2D), the three classes correspond to three different versions of 
the integer (non-interacting) quantum Hall effect.
These are the charge, spin, and thermal quantum Hall effects in classes A, C, and D;
the latter two arise in theories of 2D $d+id$ and $p+ip$ topological superconductors (TSCs). 
In three dimensions (3D), the topological classes with winding numbers $\nu \in \mathbb{Z}$ 
can describe time-reversal invariant TSCs \cite{SRFL2008}. The three TSC classes are 
distinguished by the degree of spin symmetry preserved in every quasiparticle band structure
or disorder realization; these are U(1), SU(2), and no spin symmetry for classes AIII, CI, and DIII,
respectively. Although TSCs have yet to be conclusively identified in nature, fermionic topological 
superfluids in classes A and DIII are believed to be realized in thin-film $^3$He-$A$ and 
bulk $^3$He-$B$, respectively \cite{SRFL2008,Volovik,HeliumRev}.

Topologically nontrivial phases host gapless edge or surface states at the sample boundary
\cite{KaneHasan,TSCRev1,BernevigBook} that are robust to local perturbations. In particular they should be 
protected from Anderson localization \cite{SRFL2008,Essin2015,Essin2011}. The topological protection for 
1D and 2D boundary modes is in conflict with the natural tendency of low-dimensional states to localize 
in the presence of arbitrarily weak disorder \cite{Evers2008,LeeRamakrishnan1985}. 
For 1D chiral or helical edge modes, the route of escape is that elastic backscattering is 
strictly prohibited \cite{KaneHasan,TSCRev1,KaneMele2005,Xie2016}. 
Surfaces offer a richer variety of possibilities, where topological 
bands often feature massless 2D Dirac or Majorana fermions. 
The suppression of pure backscattering for the single 2D Dirac fermion cone 
is insufficient to prevent quantum interference. Without the restriction to 1D (only forward and backward) 
in a wire, elastic impurity scattering can still occur at all other angles.
In order to resolve the puzzle in this case, it is necessary to use more technical tools like 
the nonlinear sigma model to gain further insight. For the simplest 3D topological insulator (TI), 
one finds protection of the 2D surface states from localization throughout the entire bulk energy gap.
This is understood as due to weak antilocalization enabled by strong spin-orbit coupling, 
and the presence of a $\mathbb{Z}_2$ topological term
that nullifies the metal-insulator transition in the symplectic class
\cite{Evers2008,Bardarson2007,Nomura2007,Ryu2007,Ostrovsky2007,Konig2014}. 

The 2D surface states of bulk 3D TSCs in classes AIII, CI, and DIII
are typically predicted to appear as massless Dirac or Majorana fermions. Different from graphene or TI surface states,
time-reversal invariant quenched disorder enters into these surface theories in a peculiar way.
Due to the ``fractionalization'' of the Hilbert space associated to confinement at the sample
boundary and the natural particle-hole symmetry present in a superconductor, 
2D Dirac TSC surface theories admit only quenched gauge-field disorder \cite{SRFL2008,BernardLeClair2002,Foster2014}.
In classes AIII and CI, minimal realizations involve U(1) and SU(2) vector potentials. 
The minimal realization of a class DIII surface consists of a single Majorana cone; in this case,
disorder can only modulate the velocity components of the cone. Since it couples to the stress
tensor, we call this ``quenched gravitational disorder'' (QGD) \cite{Ghorashi2020}.
Although class CI and AIII 2D Dirac models with gauge disorder could be robustly realized as TSC surface states,
these were originally studied two decades ago in the context of the high-$T_c$ cuprate superconductors 
\cite{AltlandSimonsZirnbauer2002}.
Indeed, by suppressing interpair, internode, and/or intranode elastic impurity scattering 
in a 2D $d$-wave superconductor, one can realize all three minimal surface models in classes 
CI \cite{Nersesyan1994,Tsvelik1995,Mudry1996,Caux1996,Bhaseen2001}, 
AIII \cite{Ludwig1994,Chamon1996}, and DIII \cite{Ghorashi2020}.

\begin{figure*}[t]
	\noindent \begin{centering}
	\includegraphics[width=0.85\textwidth]{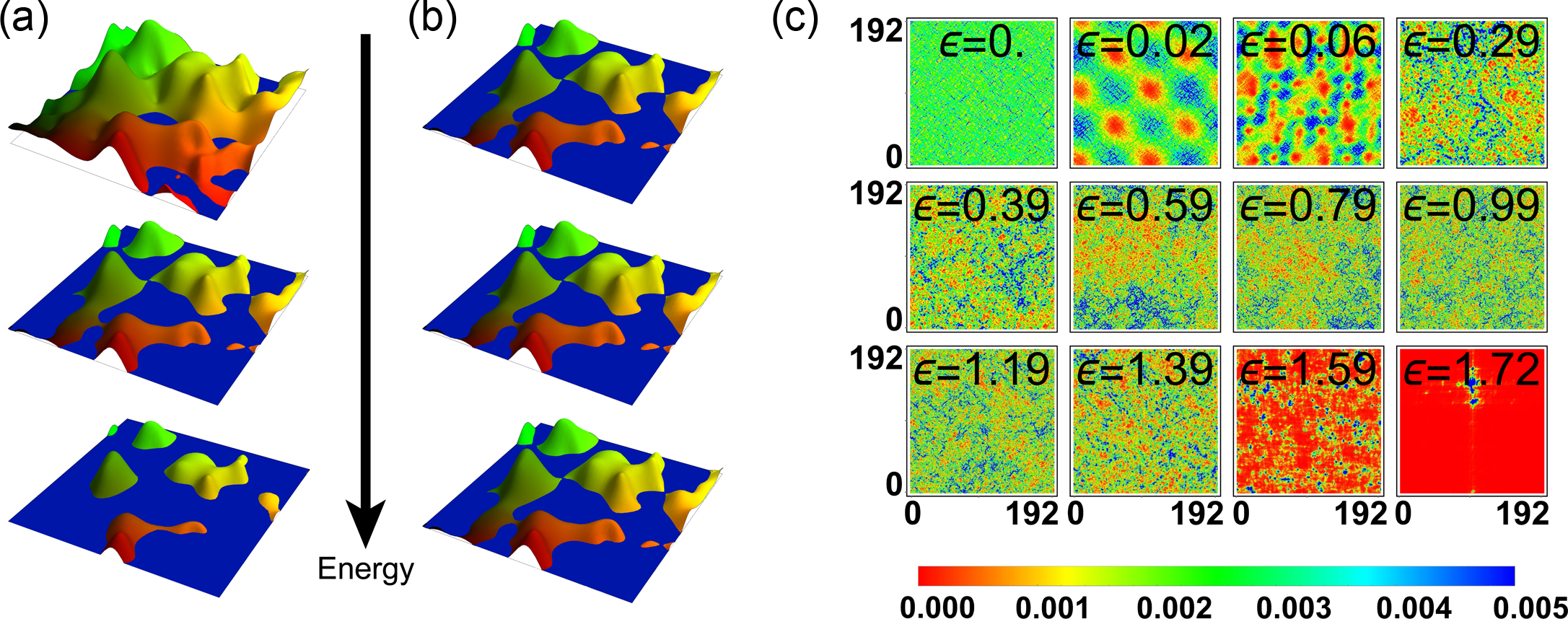}
	\par\end{centering}
	\caption{``Stacked'' quantum criticality at
	the surface of a bulk topological superconductor (TSC). 
	Panel (a) depicts classical geometric critical phenomena in 2D, as can occur
	when fluid floods a landscape. Criticality arises at the percolation
	threshold (middle), where fine-tuning of the fluid level makes
	travel across the landscape equally difficult by land or by sea. 
	By contrast, at the surface of a bulk TSC with quenched disorder
	that preserves time-reversal symmetry, the numerical studies 
	\cite{Ghorashi2020,Sbierski2020,Ghorashi2018} reviewed in this paper demonstrate a ``stacking'' of critical 
	eigenstates throughout the surface energy spectrum, schematically indicated in (b).
	Panel (c) depicts position-space probability density maps  
	for dirty TSC 2D surface eigenstates, as could be measured in the local
	density of states probed by scanning tunneling microscopy (STM). 
	Eigenstates at different energies $\varepsilon$ are shown for a 
	single class DIII surface Majorana cone, subject to a particular realization of 
	nematic quenched disorder in the components of its velocity
	(nematic ``quenched gravitational disorder'' \cite{Ghorashi2020}). 
	Eigenenergies $\varepsilon$ are measured in units of the momentum cutoff $\Lambda$
	(with average Fermi velocity set equal to one), and lengths are measured
	in units of $2\pi / \Lambda$. While low-energy states are plane-wave like 
	in this case, states with energies $0.2 < \varepsilon < 1.5$ exhibit quantum critical
	rarification. The multifractal fluctuations of the wave function intensity 
	appear to be universal, independent of $\varepsilon$ and of the disorder strength, 
	forming a ``stack'' of quantum-critical states. Importantly, evidence for 
	Anderson localization is observed only at high energies, well above the ultraviolet
	cutoff for TSC surface states in all three classes CI, AIII, and DIII;
	stacking statistics \emph{improve} for increasing system sizes and disorder strengths 
	\cite{Ghorashi2020,Sbierski2020,Ghorashi2018}.
	Results are obtained 
	by exact diagonalization of the continuum Dirac theory with periodic boundary 
	conditions, defined in momentum space (so as to avoid fermion doubling) \cite{Ghorashi2020}.  
	The ``stacked'' critical states for class AIII and CI TSC surface states 
	match the known critical statistics of the class A charge and class C spin quantum Hall
	plateau transitions, respectively \cite{Sbierski2020,Ghorashi2018}. 
	[Since the class C transition shares a few exactly known critical exponents with 2D classical percolation
	\cite{SQHPT-2_Gruzberg1999}, one can say that the stacking in class CI realizes 
	critical percolation without fine-tuning \cite{Ghorashi2018}, as sketched in panel (b).]
	The finite-energy critical fluctuations observed in class DIII, shown in (c), 
	may correspond to the thermal quantum Hall plateau transition in class D.  	
	\label{fig:stacked}}
\end{figure*}

In this paper, we review the recent numerical evidence of Refs.~\cite{Ghorashi2020,Sbierski2020,Ghorashi2018},
indicating that the class AIII, CI, and DIII Dirac surface theories evade Anderson localization 
via a highly unusual mechanism. These 2D Dirac models exhibit a ``stack'' of critical states at finite energies, 
see Fig.~\ref{fig:stacked}. 
The statistics of these states at different energies (away from zero) are identical. 
In particular, the multifractal spectrum of wave function fluctuations and the distribution
of the Landauer conductance for finite-energy class AIII Dirac surface states appear to match
the universal values associated to the quantum Hall plateau transition (QHPT) in class A \cite{Sbierski2020}. 
This is surprising for a number of reasons. First, the critical state associated to the QHPT
typically obtains only with fine-tuning of the magnetic field or particle density. This is because
the QHPT is a quantum phase transition separating topologically distinct plateaux. 
Instead, at the surface of a TSC, every finite-energy state appears to feature its own plateau transition.
Second, the quantum Hall effect lacks time-reversal symmetry (TRS), yet the findings in Ref.~\cite{Sbierski2020}
show an energy-stacking of QHPT states without TRS breaking. 
For the TSC with full spin SU(2) symmetry (class CI), the finite-energy surface states \cite{Ghorashi2018}
mimic the class C \emph{spin} QHPT phenomenology precisely \cite{Evers2008,SQHPT-1_Kagalovsky1999,SQHPT-2_Gruzberg1999,SQHPT-3_Senthil1999,SQHPT-4_Cardy2000,SQHPT-5_Beamond2002,SQHPT-6_Evers2003,SQHPT-7_Mirlin2003}. 
Finally, for the minimal realization of a class DIII surface with QGD, 
stacking occurs for a new class of wave function quantum criticality. 
This is hypothesized to be related to the thermal QHPT in class D \cite{D-1_SenthilFisher2000,D-2_BocquetZirnbauer2000,D-3_ReadLudwig2000,D-4_Gruzberg2001,D-5_Chalker2001,D-6_Mildenberger2007,D-7_Laumann2012, Fulga2020}.

The 2D Dirac surface models studied here are equivalent to Wess--Zumino--Novikov--Witten (WZNW) nonlinear sigma models \cite{BernardLeClair2002},
modified by the addition of the nonzero quasiparticle energy. The latter couples to the trace of the principal chiral
field, a strongly relevant perturbation. At zero energy (the surface Dirac point), 
these models are also quantum critical, and have been long understood thanks to the exact solution via
conformal field theory \cite{Foster2014,AltlandSimonsZirnbauer2002,Nersesyan1994,Mudry1996,Caux1996,Bhaseen2001,Ludwig1994,Guruswamy2000}.
By contrast, there is very little known analytically of the finite-energy behaviour in the perturbed WZNW models. 
Ludwig \emph{et al}.\ \cite{Ludwig1994} investigated the minimal single-node class AIII Dirac model
(corresponding to the surface of a class AIII TSC with winding number $\nu = 1$). These authors 
argued that all states at finite energy should Anderson localize. A mechanism for the finite energy states to 
escape this fate was conjectured by Ostrovsky \emph{et al}.\ \cite{Ostrovsky2007}. They showed that a gradient 
expansion yields the Pruisken model that describes the integer quantum Hall effect. For odd winding numbers $\nu$,
the Pruisken model has a theta term with topological angle $\vartheta = \pi$, corresponding to the class A QHPT. 
This result was confirmed for $\nu = 1$ by numerics \cite{Chou2014}.
While this argument supports quantum-critical stacking for odd $\nu$, it predicts Anderson localization 
for even $\nu$ (despite the $\mathbb{Z}$ classification for class AIII) \cite{Essin2015,Ostrovsky2007,Konig2014}.
There is no indication of this even/odd effect numerically \cite{Sbierski2020}, and both $\nu=1,2$ show clear indications 
of class A QHPT criticality in the multifractal spectra and conductance distribution \cite{Evers2008,Huo1993,Huckestein1995,Cho1997,Jovanovic1998,Wang1998,Evers2001,Schweitzer2005}.  Numerical evidence for class A QHPT stacking has also been very recently reported for a single Dirac cone with generic disorder~\cite{BS21}.

In this review, we summarize the key numerical findings for class AIII, CI, and DIII finite-energy surface
theories \cite{Ghorashi2020,Sbierski2020,Ghorashi2018}. For class CI, we extend the previous calculations
in Ref.~\cite{Ghorashi2018} to larger system sizes, and we provide a finite-size scaling analysis of the 
multifractal spectrum. We also present new results for the Kubo conductivity for class CI and AIII surface states. 

Beyond the connection to the hypothetical class D thermal QHPT physics, Ref.~\cite{Ghorashi2020} found that 
the single 2D Dirac or Majorana cone subject to \emph{nematic} QGD matches the phenomenology observed
in STM studies of the high-$T_c$ cuprate superconductor BSCCO 
\cite{Davis01,Davis02,Davis05-a,Davis05-b,Davis08,DavisReview}, see Sec.~\ref{sec:QGD}.
The field of experimental studies of disordered superconductors is very rich by itself. 
Many theoretical scenarios for the disorder-driven superconductor-to-insulator transition  
involve enhanced Cooper pairing, due to multifractal rarification 
\cite{Feigelman07,Feigelman10,Foster12,Foster14,Burmistrov15}. 
Reporting an increase in $T_c$ with increasing disorder, Ref.~\cite{Welp18} recently added experimental 
support for this. Furthermore there are indications that multifractal superconductor physics provides 
an adequate description of transition metal dichalcogenides \cite{Ji19,Ugeda18,Yeom19}. 
It is also interesting to note that the chiral model for twisted bilayer graphene is effectively 
described by a class CI surface Dirac Hamiltonian \cite{Bistritzer2011,Guinea2012,Vishwanath2019}.
However, the most prominent and well-studied experiments revolve around the mystery of the spatial 
inhomogeneity in the high-$T_c$ cuprate superconductors \cite{DavisReview}; the results reviewed here call for a
re-evaluation of the role of disorder in these materials.

\subsection{Outline}

In this paper, we first give a brief overview about topological surface theories. We start with the topological classification and the corresponding $\sigma$ models (see Table~\ref{tab:10-fold-way}), including the conventional expectations for the finite-energy behaviour. Next we review the key results of WZNW theory relevant for zero-energy states of time-reversal invariant superconducting classes. We explain the role of the energy perturbation and how the modified WZNW model can be deformed ``by hand'' into the Pruisken model.

The main part is organized as a review of the most important numerical results for the class AIII, CI, and DIII surface theories. Each section about these theories covers the multifractal spectra, and for class AIII the Landauer conductance distribution.
New results in this review include Kubo conductivity computations for classes AIII and CI, as well as larger system sizes and a finite-size analysis for class CI, winding number $\nu=2$ multifractal spectra.


\section{Fundamentals \label{sec:Fund}}

\subsection{Dirac surface theories and topological classification \label{sec:topclass}}

We consider 2D surface theories of 3D bulk, time reversal ($\mathcal{T}$)-invariant topological superconductors (TSCs) with different degrees of spin symmetry. These reside in classes AIII, CI, or DIII, as indicated in Table~\ref{tab:10-fold-way}. For any superconducting realization of a class, \emph{physical} time-reversal symmetry $\mathcal{T}$ corresponds to the \emph{effective} chiral symmetry $S$ in this table. This symmetry transmutation is due to the ``automatic'' particle-hole invariance associated to the self-conjugate (Balian-Werthammer) spinor formulation of any Bogoliubov-de Gennes Hamiltonian  \cite{SRFL2008,Foster2008}. 
These bulk phases can be topologically non-trivial, and are indexed with integer-valued winding numbers.

\begin{table*}
	{\renewcommand{\arraystretch}{1.2}
	\arraycolsep=1.4pt\def\arraystretch{2.2}
		\begin{tabular*}{\textwidth}{l @{\extracolsep{\fill}} cccccclc}
			\hline
			\hline
			Class \,\,
			& 
			$T$
			&
			$P$
			& 
			$S$
			&
			Spin sym. 
			&
			$d = 2$
			&
			$d = 3$
			& 
			Realization	
			& 
			Fermion NL$\sigma$M	
			\\
			\hline
			C 				& 0 & -1 & 0 	& SU(2) & $2 \mathbb{Z}$ 	& - 			& SQHE		&	 $\dfrac{\mathrm{Sp}(4n)} {\mathrm{U}(2n)}$
			\\
			A 		& 0 & 0 & 0 	& U(1) 	& $\mathbb{Z}$ 		& - 			& IQHE 				&	$\dfrac{\mathrm{U}(2n)} {\mathrm{U}(n) \otimes \mathrm{U}(n)}$
			\\
			D 				& 0 & +1 & 0 	& - 	& $\mathbb{Z}$ 		& - 			& TQHE	&	$\dfrac{\mathrm{O}(2n)} {\mathrm{U}(n)}$ 
			\\
			CI 				& +1 & -1 & 1 	& SU(2) & -		 	& $2 \mathbb{Z}$ 	& 3D TSC 			& 	$\dfrac{\mathrm{Sp}(4n) \otimes \mathrm{Sp}(4n)} {\mathrm{Sp}(4n)}$
			\\
			AIII 				& 0 & 0 & 1	& U(1) 	& - 			& $\mathbb{Z}$ 	 	& 3D TSC, chiral TI		& 	$\dfrac{\mathrm{U}(2n) \otimes \mathrm{U}(2n)} {\mathrm{U}(2n)}$
			\\
			DIII 				& -1 & +1 & 1 	& - 	& $\mathbb{Z}_2$	& $\mathbb{Z}$ 	& 3D TSC ($^3$He-$B$)			& 	$\dfrac{\mathrm{O}(2n) \otimes \mathrm{O}(2n)} {\mathrm{O}(2n)}$ 
			\\
			AI            & +1 & 0 & 0 	& SU(2) & -		 	& - 			& -		 		& 	$\dfrac{\mathrm{Sp}(4n)} {\mathrm{Sp}(2n) \otimes \mathrm{Sp}(2n)}$
			\\
			AII 	      & -1 & 0 & 0	& - 	& $\mathbb{Z}_2$ 	& $\mathbb{Z}_2$ 	& 2D, 3D TIs 			& 	$\dfrac{\mathrm{O}(2n) } {\mathrm{O}(n) \otimes \mathrm{O}(n)}$
			\\
			BDI 				& +1 & +1 & 1 	& SU(2)	& -		 	& - 			& -		 		& 	$\dfrac{\mathrm{U}(2n)} {\mathrm{Sp}(2n)}$
			\\
			CII 			 	& -1 & -1 & 1	& - 	& -		 	& $\mathbb{Z}_2$ 	& 3D chiral TI 			& 	$\dfrac{\mathrm{U}(2n) } {\mathrm{O}(2n)}$
			\\
			\hline 
			\hline
		\end{tabular*}
	}
	\caption{The 10-fold way classification for strong (fully gapped), 
			$d$-dimensional symmetry-protected topological phases of 
			fermions, i.e.\ 
			topological insulators (TIs) and topological superconductors (TSCs) \cite{SRFL2008,Kitaev2009,SRFL2010,TSCRev2}. 
			The 10 classes are defined by different combinations of the three \emph{effective} discrete symmetries 
			$T$ (time--reversal), 
			$P$ (particle--hole), 
			and 
			$S$ (chiral or sublattice).
			For a $d$-dimensional bulk, any deformation of the clean band structure that preserves $T$, $P$, and $S$ 
			and does not close a gap preserves the topological winding number.
			For a $(d-1)$-dimensional edge or surface theory, the equivalent statement is that any static
			deformation of the surface (quenched disorder) that preserves $T$, $P$, and $S$ also preserves the 
			``topological protection'' against Anderson localization. 
			Of particular interest here are classes C, A, D on one hand, and classes CI, AIII, DIII on the other. 
			Classes C, A, and D are topological in $d = 2$, and describe the spin (SQHE), integer or charge (IQHE), and thermal (TQHE) quantum Hall effects;
			all three can be realized as TSCs with broken $T$. 
			Classes CI, AIII, and DIII are topological in $d = 3$, and can describe 3D time-reversal-invariant TSCs. 
			(In this case, the physical time-reversal symmetry appears as the \emph{effective} chiral symmetry $S$ in the table \cite{SRFL2008,Foster2008}.)
			The column ``spin sym.'' denotes the amount of spin SU(2) symmetry preserved for TSC realizations of these 6 classes. 
			The 3D TSCs can host 2D massless Dirac (CI, AIII) or Majorana (DIII) surface theories. 
			In the presence of disorder that preserves physical time-reversal symmetry, these are equivalent to Wess--Zumino--Novikov--Witten (WZNW) sigma models, modified by a relevant perturbation
			for nonzero surface eigenenergy $\varepsilon$. While the $\varepsilon = 0$ WZNW models have long been understood 
			to exhibit critical delocalization
			\cite{Foster2014,AltlandSimonsZirnbauer2002,Nersesyan1994,Mudry1996,Caux1996,Bhaseen2001,Ludwig1994,Guruswamy2000}, 
			the conventional expectation was that $\varepsilon \neq 0$ breaks the defining chiral $S$ symmetry,
			producing a standard Wigner-Dyson class,
			so that \cite{Evers2008}
			CI 	$\rightarrow$ AI 	(Anderson localized),
			AIII 	$\rightarrow$ A 	(Anderson localized),
			and
			DIII 	$\rightarrow$ AII	(Anderson localized or weak antilocalization).
			The numerical results of \cite{Sbierski2020,Ghorashi2018} instead establish that the 2D Dirac surface
			theories in classes CI and AIII exhibit ``critical stacking'' (see text) of quantum Hall plateau
			transition states in classes C and A, respectively.
			Critical stacking for class DIII is conjectured to correspond to the class D TQHE \cite{Ghorashi2020}.
			The last column gives the symmetry structure of the non-linear sigma model (NL$\sigma$M) description for each 
			class, in terms of fermionic replicas \cite{Evers2008}. 
	}
	\label{tab:10-fold-way}
\end{table*}

The form of the 2D surface band structure for a clean topological phase in general depends upon some details of the bulk
and of the surface orientation.
A large class of TSC surface states in classes AIII, CI, and DIII take the form of massless Dirac
or Majorana fermions. This has been demonstrated using bulk lattice models in (e.g.) Refs.~\cite{Schnyder2009,Hosur2010,Schnyder2010,Xie2015}.
Generic $\mathcal{T}$-invariant quenched disorder introduced at the surface translates into 
random abelian and/or nonabelian vector potentials in the low-energy surface Dirac theory.
A generic Hamiltonian is \cite{SRFL2008,Foster2014}
\begin{align}
	H 
	&= 
	{\sigb}
	\cdot
	\left[
		(-i \Nabla) + \sum_{i}{\vex{A}_i}({\vex{r}}) \, {\tauh}^i
	\right] 
	\label{eq:surface}.
\end{align}
Although this is a single--particle Hamiltonian for (2+1)-D surface quasiparticles,
it is frequently useful to alternatively interpret $H$ as the Lagrangian density for
an imaginary time (2+0)-D theory of 1D relativistic fermions. 
The pseudospin Pauli matrices $\sigh^{1,2}$ then act separately on the spaces of left- and right-movers \cite{Foster2014}. 
The matrices $\{\tauh^i\}$ act upon an $N$-dimensional color space, and couple to the nonabelian vector potential $\vex{A}_i$. The color generators have to be compatible with the symmetry of the class. In class AIII, for a bulk winding number $\nu = N$, there can be generic $\text{U}(N)$ disorder that encodes elastic scattering between the colors. 
Thus all Hermitian $N \times N$ generators $\{\tauh^i\}$ are allowed, including the identity matrix [U(1) abelian vector potential disorder].
For class DIII, these are restricted to antisymmetric generators of SO($N$). 
In class CI, the winding number $\nu = N \equiv 2 M$ is always even, and the matrices $\{\tauh^i\}$ generate the Lie algebra Sp($2M$). 

The key defining characteristic of a topological surface is the anomalous representation of a defining bulk symmetry. 
For surface states of 3D TSCs, this is the chiral/physical time-reversal symmetry. For the Hamiltonian in Eq.~(\ref{eq:surface}),
it is encoded by the condition
\begin{align}\label{eq:chiral}
	\sigh^3 \, H + H \, \sigh^3 = 0.
\end{align}
This version of chiral symmetry is anomalous, i.e.\ cannot arise without fine-tuning from the continuum Dirac description of
a 2D lattice model \cite{SRFL2008,Foster2014}. It can be shown that Eq.~(\ref{eq:chiral}) 
implies that the class CI, AIII, or DIII nonlinear sigma model (NL$\sigma$M) encoding Anderson (de)localization physics \cite{Evers2008}
is augmented by a Wess--Zumino--Novikov--Witten (WZNW) term \cite{SRFL2008,BernardLeClair2002}. Without the WZNW
terms, the NL$\sigma$Ms in these classes are termed ``principal chiral models'' or principal chiral NL$\sigma$Ms.   
The minimal realizations of Eq.~(\ref{eq:surface}) for topological class CI, AIII, and DIII surfaces 
have winding numbers $\nu = N = \{2,1,1\}$, respectively. 

By contrast, the minimal ``non-topological'' version of class CI possesses four colors of 
2D Dirac fermions \cite{AltlandSimonsZirnbauer2002}. Incorporating disorder, the generic continuum Dirac model corresponding to a dirty 2D $d$-wave superconductor is perturbed by random mass, vector, and scalar potential terms. This model is believed to Anderson localize for arbitrarily weak disorder at all energies; it corresponds to the class CI principal chiral nonlinear sigma model \emph{without} 
the WZNW term \cite{AltlandSimonsZirnbauer2002,Senthil1998}. 
At the same time, the $d$-wave model can be fine-tuned to realize any of the three topological models as exemplified by Eq.~(\ref{eq:surface}) \cite{Ghorashi2020,AltlandSimonsZirnbauer2002}.
Suppressing elastic scattering between \emph{pairs} of nodes gives two copies of the $\nu = 2$ class CI WZNW model (nodes in a pair are related by $\mathcal{T}$). Further suppressing scattering between nodes within a pair breaks each $\nu = 2$ CI model into two $\nu = 1$ AIII WZNW models,
with only U(1) vector disorder. Suppressing even this still allows random fluctuations of the velocity components, which correspond to ``quenched gravitational disorder'' in class DIII \cite{Ghorashi2020}. 
This example displays a general rule: a non-topological class CI, AIII, or DIII model (associated e.g.\ to a 2D lattice model) can always be fragmented into topological components, provided restrictions are placed upon scattering between the different Dirac colors. 
These restrictions cannot, however, typically be realized exactly in a microscopic 2D model with lattice-scale disorder.

The averages over ensembles of disordered $H$ in Eq.~(\ref{eq:surface}) can be described by the NL$\sigma$M theory \cite{Evers2008}. 
Using fermionic replicas, the topological surface-state WZNW Dirac models are associated to the group manifolds 
$G(2n) \in \{\text{U}(2n),\text{Sp}(4n),\text{O}(2n)\}$ for classes AIII, CI, and DIII, respectively, as shown in Table~\ref{tab:10-fold-way}.
Here $n \rightarrow 0$ denotes the number of replicas. 
The conventional expectation is that any non-standard symmetry class such as these must reduce to a standard
Wigner--Dyson class (A, AI, or AII) at finite eigenstate energy $\varepsilon \neq 0$. This is because $\varepsilon \neq 0$ formally 
breaks the defining particle--hole or chiral symmetry. Classes AIII, CI, and DIII exhibit $G \otimes G$ symmetry at zero
energy, but this is reduced to the diagonal subgroup $G$ for $\varepsilon \neq 0$ (see also Sec.~\ref{sec:wzw}).
One would therefore expect that finite-energy states in classes AIII, CI, and DIII reside in classes A, AI, and AII, respectively,
characterized by the symmetry reduction in the NL$\sigma$M from the group manifolds to the corresponding Grassmannians, 
\begin{align}
	\frac{G(2n) \otimes G(2n)}{G(2n)}
	\simeq
	G(2 n) 
	\rightarrow 
	\frac{G(2n)}{G(n) \otimes G(n)},
\end{align}
see Table~\ref{tab:10-fold-way}.
Class AI always localizes in 2D, as does class A unless fine-tuned to the QHPT; class AII can exhibit weak antilocalization
for sufficiently weak disorder \cite{Evers2008,LeeRamakrishnan1985}.

Although we limit our focus in this review to Dirac surface theories, 
there are other possibilities. Bulk TSCs or fermionic topological superfluids that arise by 
pairing higher-spin fermions (e.g., $S = 3/2$) can give rise to surface Hamiltonians
that also exhibit the anomalous chiral symmetry in Eq.~(\ref{eq:chiral}). These have 
larger minimal winding numbers $|\nu| > 1$, and the bulk winding number can be reflected
through nonlinearity of the surface band structure, instead of $N = |\nu|$ colors
of linearly-dispersing Dirac fermions \cite{Ghorashi2018,Wu2016,Fang2015,Ghorashi2017,Roy2019}.
Numerical studies suggest that the disorder-induced physics of these surfaces
are the same as the Dirac models studied here \cite{Ghorashi2018,Ghorashi2017,Roy2019}.


\subsection{Multifractality}

Critical disordered systems can be characterized by the scaling of the distribution of 
powers of the wave function $|\psi|^2$, the so called inverse-participation ratios $P_q$:
\begin{align}\label{eq:ipr}
	P_q 
	&\equiv 
	\int d^2 \vex{r} 
	\, |\psi(\vex{r})|^{2q} 
	\sim 
	L^{-\tau_q}.
\end{align}
In the limit that the system-size $L \rightarrow \infty$, the multifractal exponents $\tau_q$ are self-averaging \cite{Evers2008}. 
Numerically it is favorable to subdivide a $L \times L$ system into $N^2$ boxes of size $b$, with $N \equiv L/b$. 
The box probability $\mu_i \equiv \intl{b_i} d^2\vex{r} \, |\psi_0(\vex{r})|^2$ shows scaling behavior suitable to extract $\tau_q$:
\begin{align}\label{tau(q)Def}
	\sum_{i = 1}^{N^2} (\mu_i)^q \sim \left(\frac{b}{L}\right)^{\tau_q}. 
\end{align}
This way one can handle correlated disorder more easily, by excluding boxes smaller 
than the disorder correlation length.

The anomalous dimensions
\begin{align}\label{Delta(q)Def}
	\Delta_q
	\equiv 
	\tau_q - 2(q - 1)
\end{align}
can be interpreted as the deviation from the scaling of a fully delocalized metallic wave function.
Localized states instead have $\tau_q = 0$ for $q > 0$. 

In subsequent sections, we extract the multifractal spectra of the Dirac Hamiltonians that we consider 
and compare these to universal predictions for two classes of theories, 
which now review in Secs.~\ref{sec:wzw} and \ref{sec:qh}.


\subsection{Class AIII, CI, and DIII WZNW models over $\text{U}(2n)$, $\text{Sp}(4n)$, and $\text{O}(2n)$}\label{sec:wzw}

The statistics of the spatial fluctuations for eigenstates of 2D disordered systems are described 
by the non-linear sigma model (NL$\sigma$M) framework \cite{Evers2008}. Specifically, for the class AIII, CI, 
or DIII topological surface-state Hamiltonian in Eq.~(\ref{eq:surface}), this sigma model 
becomes a Wess--Zumino--Novikov--Witten (WZNW) model, familiar from conformal field theory (CFT). 
Using non-abelian bosonization 
\cite{AltlandSimonsZirnbauer2002,Nersesyan1994,Mudry1996,Caux1996,Bhaseen2001,Ludwig1994,Guruswamy2000} 
and conformal embedding theory \cite{Foster2014}, one can derive exact results for the scaling of 
generic operators in the energy $\varepsilon \rightarrow 0$ limit. 
In particular, for the density of states (DOS) $\rho(\e)$ as a function of the surface quasiparticle 
energy $\e$, one has 
\begin{align}\label{DoS_nu}
	\lim_{\e \rightarrow 0} 
	\rho(\e) 
	\simeq 
	|\e|^{x_1/z}. 
\end{align}
Here $x_1 = 2 - z$ is the scaling dimension of the operator encoding the first moment of the local 
density of states at $\e = 0$, and $z$ denotes the dynamic critical exponent. 
For surface states of a bulk TSC with winding number $\nu$, the scaling exponent $x_1/z$ is 
summarized for the different WZNW models in Table~\ref{tab:WZW}.
The multifractal spectrum for the WZNW models at $\e = 0$ is exactly parabolic:
\begin{align}\label{MFC_WZNW}
	\Delta_q &= \theta \, q(1-q).
\end{align}
Table~\ref{tab:WZW} summarizes how $\theta$ depends upon the class and winding number.

The WZNW action for 2D dirty Dirac or Majorana TSC surface states with winding number $\nu$ 
reads 
\begin{align}
\!\!\!
S
=&\,
\frac{\nu}{8 \pi l_\phi}
\int
d^2\vex{r}
\,
\Tr\!\left[
\Nabla \hat{Q}^\dagger 
\cdot
\Nabla \hat{Q} 
\right]
-
\frac{i \nu}{12 \pi l_\phi}
\int
d^2\vex{r}
\,
d R
\,
\epsilon^{a b c}
\Tr\!\left[
\left(\hat{Q}^\dagger \partial_a \hat{Q}\right)
\left(\hat{Q}^\dagger \partial_b \hat{Q}\right)
\left(\hat{Q}^\dagger \partial_c \hat{Q}\right)
\right]
\nonumber\\&\,
-
\frac{\lambda_A \, \nu^2}{8 \pi^2}
\int
d^2\vex{r}
\,
\Tr\!\left[
\hat{Q}^\dagger
\Nabla 
\hat{Q}
\right]
\cdot
\Tr\!\left[
\hat{Q}^\dagger
\Nabla 
\hat{Q}
\right]
+
\frac{i \omega}{2}
\int
d^2\vex{r}
\,
\Tr\!\left[
\hat{\Lambda}
\left(
\hat{Q} + \hat{Q}^\dagger
\right)
\right].
\label{WZNW}
\end{align}
See e.g.\ Ref.~\cite{Foster2014} for a derivation of this action
from the disordered Dirac theory defined by Eq.~(\ref{eq:surface}).
The zero-energy surface theory for classes CI and AIII
is described by the top line equation~(\ref{WZNW}). The WZNW term is the second one
on this top line, and requires extending the field configurations from the 2D surface
into the 3D bulk \cite{Foster2014,Koenig2012}; the parameter $l_\phi$ is the Dynkin index
of the corresponding group. 
For class AIII only, an additional parameter appears even at zero energy,
which is the marginal disorder strength $\lambda_A$ that encodes the strength of abelian vector potential disorder. 

The parameter $\omega$ on the second line of Eq.~(\ref{WZNW}) is the ac frequency at which the conductivity 
of the NL$\sigma$M is to be evaluated. With $\omega\neq 0$, states at finite energy can be accessed. 
This parameter couples to the \emph{imaginary} (``tachyonic'')  mass term 
$(i/2) \Tr[\hat{\Lambda}(\hat{Q} + \hat{Q}^\dagger)]$, where $\hat{\Lambda} = \diag{\{\hat{1}_n,-\hat{1}_n\}}$ 
grades in the retarded/advanced space \cite{Evers2008}. 
Since it is a strongly relevant perturbation, nonzero $\omega$ drives the theory away from quantum critical
point solved by the WZNW conformal field theory. 

The field $\hat{Q}(\vex{r})$ is a $(2 n)$$\times$$(2 n)$ [$(4 n)$$\times$$(4 n)$] element of the matrix group
$\text{U}(2n)$, $\text{O}(2n)$, [$\text{Sp}(4n)$] for classes AIII, DIII [CI]. 
It satisfies the nonlinear constraint 
$\hat{Q}^\dagger(\vex{r}) \, \hat{Q}(\vex{r}) = \hat{1}$, where $\hat{1}$ denotes the identity matrix. 
In the end, the replica limit $n \rightarrow 0$ has to be taken \cite{Evers2008,Foster2014}. 

For $\omega = 0$, the WZNW model in Eq.~(\ref{WZNW}) is exactly solvable via CFT
\cite{Foster2014,Nersesyan1994,Tsvelik1995,Mudry1996,Caux1996,Bhaseen2001,Ludwig1994,Guruswamy2000,Ostrovsky2006}.
Exact results for the DOS scaling [Eq.~(\ref{DoS_nu})], multifractal spectrum [Eq.~(\ref{MFC_WZNW})], 
and conductivity are summarized in Table~\ref{tab:WZW}.

\begin{table*}[t!]
	\def\arraystretch{1.5}
	\begin{tabular}{c|ccc}
		\hline 
		\hline
		& AIII & CI & DIII\\
		\hline
		\hline
		$\nu$ & $\mathbb{Z}$ & $2\mathbb{Z}$ & $\mathbb{Z}$ 
		\\
		$x_1/z$ 
		& 
			${\frac{\pi - \nu^2 \lambda_A}{\pi(2 \nu^2-1) + \nu^2 \lambda_A}}$ \cite{Ludwig1994} 
		& 
			$\frac{1}{2|\nu|+3}$ \cite{Nersesyan1994} 
		& 
			$-\frac{1}{2|\nu|-3}$ \, ($|\nu| \geq 3$) \cite{Foster2014,LeClair2008}
		\\
		$\theta(\e = 0)$ 
		& 
			$\frac{|\nu| - 1}{\nu^2}+ \frac{\lambda_A}{\pi}$ \cite{Mudry1996,Caux1996,Ludwig1994} 
		& 
			$\frac{1}{|\nu|+2}$ \cite{Mudry1996,Caux1996,Foster12}  
		&  
			$\frac{1}{|\nu|-2}$ \, ($|\nu| \geq 3$) \cite{Foster2014} 
		\\
		$\sigma^{xx}(\e = 0)$
		& 
			$\frac{\nu}{\pi} $ 
		& 
			$\frac{\nu}{\pi} $ 
		& 
			$\frac{\nu}{\pi} $ 
		\\
		\hline
		$\theta(\e \neq 0)$ 
		& 
			$\simeq 1/4$ (IQHPT) \cite{Sbierski2020,Chou2014} 
		& 
			$\simeq 1/8$ (SQHPT) \cite{Ghorashi2018} 
		& 
			$\simeq 1/13$ (TQHPT?) \cite{Ghorashi2020}  
		\\
		$\sigma^{xx}(\e \neq 0)$ 
		& 
			$\simeq 0.58 \pm 0.02$ \cite{Schweitzer2005} 
		& 
			$= \frac{\sqrt{3}}{2}$ \cite{SQHPT-4_Cardy2000} 
		& ? 
		\\
		\hline 
		\hline
	\end{tabular}
	\caption{Summary of known properties
	for the 2D disordered Dirac models [Eq.~(\ref{eq:surface})] 
	in classes AIII, CI, DIII that can exhibit Wess--Zumino--Novikov--Witten (``stacked'') criticality 
	at zero (nonzero) energy.
	Here these are 
	the allowed bulk TSC winding numbers $\nu$, 
	the scaling of the surface density of states $\rho(\e)\propto |\e|^{x_1/z}$, 
	the curvature of the parabola $\theta$ controlling the surface multifractal spectrum via 
	$\Delta_q = -\theta \, q(1-q)$, 
	and the longitudinal surface conductivity $\sigma^{xx}$
	(for spin or heat transport at the boundary of the TSC, in units of the appropriate conductance quantum \cite{SRFL2008,Xie2015}). 
	The top four rows describe the zero or near-zero energy critical features of 
	dirty 2D TSC surface states, which are known analytically from conformal field theory.   
	In class AIII, these results depend on the winding number $\nu$ 
	and the abelian disorder strength $\lambda_A$, which is defined in 
	Eqs.~\eqref{AIII-1} and \eqref{AIII-2}. The additional parameter $\lambda_A$ 
	is RG-marginal and addresses a continuum of distinct zero-energy fixed points \cite{Mudry1996,Caux1996,Ludwig1994}. 
	The last two rows detail the recent numerical findings of 
	Refs.~\cite{Sbierski2020}, \cite{Ghorashi2020}, and \cite{Ghorashi2018}. 
	These characterize the ``stacked'' criticality of TSC surface states at finite energy,
	where each state in the stack exhibits identical statistical properties. 
	The results for the finite-energy multifractal spectra and conductance statistics 
	are consistent with a stacking of the integer class A and spin class C quantum Hall plateau
	transition (QHPT) states for class AIII and CI Dirac models, respectively. 
	The stacked criticality observed for finite-energy class DIII states
	is conjectured to describe the thermal QHPT in class D \cite{Ghorashi2020}.
	We note that the numerical results for finite-energy class DIII states 
	have only been obtained for quenched gravitational disorder, 
	i.e.\ modulation of the velocity components for the single Majorana cone
	associated to winding number $\nu = 1$. 
	The multicolor DIII model with $|\nu| = N > 2$ colors has not been studied numerically
	(but see Refs.~\cite{Ghorashi2017,Roy2019}).
	}~\label{tab:WZW}
\end{table*}

\paragraph{Finite energy behavior}
The ac frequency parameter $\omega$ in Eq.~(\ref{WZNW}) reduces
the $G \otimes G$ group symmetry of the WZNW model down to the diagonal subgroup $G$. 
Real nonzero $\omega$ gives oscillatory contributions to the functional 
integral over $\hat{Q}$ unless a further constraint is imposed, 
\begin{align}\label{Constraint}
	\hat{Q} = \hat{Q}^\dagger,
	\quad
	\Tr\left[\hat{Q}\right] = 0
\end{align}
(after absorbing the matrix $\hat{\Lambda}$ by a left-group translation $\hat{\Lambda} \hat{Q} \rightarrow \hat{Q}$). 
Then, the $\omega$ term and the $\lambda_A$ term (class AIII) on the second line of Eq.~(\ref{WZNW}) 
are projected to zero. 

In this constrained case, 
Bocquet, Serban, and Zirnbauer \cite{D-2_BocquetZirnbauer2000} 
(see also \cite{Konig2014,AltlandSimonsZirnbauer2002}) derived a deformation 
of the WZNW term to the topological term in the Pruisken model:
\begin{align}
S
\rightarrow
&\,
\frac{\sigma_{x,x}}{8}
\int
d^2\vex{r}
\,
\Tr\!\left[
\Nabla \hat{Q}
\cdot
\Nabla \hat{Q} 
\right]
-
\frac{\sigma_{x,y}}{8}
\int
d^2\vex{r}
\,
\epsilon^{i j}
\Tr\!\left[
\hat{Q} 
\, 
\partial_i \hat{Q}
\,
\partial_j \hat{Q}
\right]\!,
\label{Pruisken}
\end{align}
where 
\begin{align}\label{PruiskenParams}
\sigma_{x,x} = \nu/\pi,
\quad
\sigma_{x,y} = \nu/2.
\end{align}

For classes CI and DIII, the Pruisken model only applies
if the target manifold for the constrained $\hat{Q}$ is taken 
to be that of classes C and D, respectively. 
Although $\omega \neq 0$ reduces the symmetry of the WZNW action 
down to the diagonal $G$ subgroup, this information is insufficient 
to determine the target manifold $G/H$ of the effective NL$\sigma$M
governing the Anderson (de)localization physics of the finite-energy 
states. In the case of class CI with $G = \text{Sp}(4n)$ (using
fermionic replicas, Table~\ref{tab:10-fold-way}), 
there are two possible scenarios for the finite-energy NL$\sigma$M.
Either $H = \text{Sp}(2n)$ $\otimes$ $\text{Sp}(2n)$ (the orthogonal Wigner-Dyson
class AI), or $H = \text{U}(2n)$ (class C). 
The former choice is the conventional one that guarantees
Anderson localization at all finite energies \cite{Evers2008}; the latter
is realized in the stacking scenario, wherein Eq.~(\ref{Pruisken})
describes the \emph{spin} quantum Hall plateau transition
\cite{Evers2008,SQHPT-1_Kagalovsky1999,SQHPT-2_Gruzberg1999,SQHPT-3_Senthil1999,SQHPT-4_Cardy2000,SQHPT-5_Beamond2002,SQHPT-6_Evers2003,SQHPT-7_Mirlin2003}.

Although there is no ambiguity in the target manifold for finite-energy 
class AIII states (which reside in class A), 
the ``derivation'' of Eq.~(\ref{Pruisken}) 
from Eq.~(\ref{WZNW}) via the imposition of the constraint in Eq.~(\ref{Constraint})
poses another problem. 
Eq.~(\ref{PruiskenParams}) implies that the topological angle 
$\vartheta = 2 \pi \sigma_{x,y}$ is an odd (even) multiple of $2 \pi$ 
for odd (even) winding numbers $\nu$. 
This even-odd effect is \emph{not} observed in the numerics here and in Refs.~\cite{Sbierski2020,Ghorashi2018}. 
In other words, imposing Eq.~(\ref{Constraint}) by hand directly to the fields gives 
coefficients of the Pruisken model that are not compatible with numerics. 
This does not rule out this analytical ansatz as a description of the problem, 
as the following explanation clarifies. The actual physical RG flow of the full WZNW theory 
is that $\omega$ runs to the strong coupling regime. 
The other coefficients are likely to receive renormalization well before the 
$\hat{Q}$ field is reduced to the target manifold associated to Eq.~(\ref{Constraint}). 
Consequently, the physical Pruisken model parameters can deviate 
from the values stated in Eq.~(\ref{PruiskenParams}).


\subsection{Classes A, C, and D quantum Hall criticality}
\label{sec:qh}

We claim that time-reversal invariant 3D TSCs show stacks of quantum Hall plateau transition (QHPT) states at finite energy. 
Therefore, a brief recapitulation of the features in QH systems that we compare with our numerics is appropriate.
\paragraph{Integer Quantum Hall Plateau Transition --- class A}
Studies of the network model \cite{Evers2001} indicate parabolic multifractality with $\theta \simeq 1/4$ 
in Eq.~(\ref{MFC_WZNW}).
The longitudinal conductivity is known to be $\sigma^{xx}_{\mathrm{IQHPT}} \simeq 0.58\pm 0.02$ \cite{Schweitzer2005} from numerical 
Kubo computations. Excitingly, the IQHPT transition is conjectured to itself be described by a class AIII $n=4$ WZNW CFT 
\cite{Zirnbauer2019,DSG2021}. If true, this would correspond to a generalized version of Haldane's conjecture for one-dimensional 
half-integer spin quantum antiferromagnets. It would imply a \emph{reverse} RG flow from the Pruisken model 
[Eq.~(\ref{Pruisken})] to the class AIII WZNW action [Eq.~(\ref{WZNW}) with $\nu = 4$, $\lambda_A = \pi/16$, and $\omega = 0$ 
\cite{Sbierski2020,Zirnbauer2019}]. Moreover, this indicates that class AIII $\nu=4$ TSC surfaces might be described 
by the \emph{same} conformal field theory (albeit with a renormalized value of $\lambda_A$) at both zero and finite energies \cite{Sbierski2020}.

\paragraph{Spin Quantum Hall Plateau Transition --- class C}
The SQHPT transition is well-studied analytically by a mapping to 2D classical percolation, and numerically with the network model \cite{SQHPT-1_Kagalovsky1999, SQHPT-2_Gruzberg1999, SQHPT-3_Senthil1999, SQHPT-4_Cardy2000, SQHPT-5_Beamond2002, SQHPT-6_Evers2003, SQHPT-7_Mirlin2003}. The conductance distribution and exact average longitudinal conductivity 
$\sigma^{xx}_{\mathrm{SQHPT}}=\sqrt{3}/2$ 
\cite{SQHPT-4_Cardy2000} have been determined. For certain multifractal exponents there are exact analytical results available compatible with $\theta=1/8$ in Eq.~(\ref{MFC_WZNW}) (assuming parabolicity).

\paragraph{Thermal Quantum Hall Plateau Transition --- class D}
Class D permits a variety of distinct network models and Hamiltonians \cite{D-1_SenthilFisher2000, D-2_BocquetZirnbauer2000, D-3_ReadLudwig2000, D-4_Gruzberg2001, D-5_Chalker2001, D-5_Chalker2001, D-6_Mildenberger2007, D-7_Laumann2012}, and there is non-universality. The conjectured thermal QHPT transition is difficult to observe, since it may be shaded by a weakly antilocalizing thermal metal phase.


\section{Axial U(1) spin symmetry: class AIII}
\label{sec:aiii}

In this section we review the numerical results from Ref.~\cite{Sbierski2020} for the 
2D Dirac TSC surface theory in class AIII, and the AIII WZNW~$\rightarrow$~A IQHPT stacking conjecture. 
As pointed out already in Sec.~\ref{sec:wzw}, it is important to distinguish between even and odd winding numbers. 
Computationally it is easiest to look at $\nu=1,2$.

\paragraph{One Dirac node, U(1) vector potential dirt (AIII, $\nu=1$)\label{sec:nu=1}}
The winding number $\nu=1$ AIII surface theory can be realized with a single Dirac cone (addressed with Pauli matrices $\sigh^{1,2}$).
\begin{align}\label{AIII-1}
	H_{\mathrm{AIII}}^{\pup{1}} 
	=
	\sigb\cdot\left[-i \, \vv \, \Nabla + \vex{A}(\vex{r})\right],
	\qquad
	\overline{A^a(\vex{r}) \, A^{b}(\vex{r'})} 
	=
	\lambda_A \, \delta^{a b} \, \delta_\xi^\pup{2}(\vex{r} - \vex{r'}).
\end{align}
Here $\vv$ is the Fermi velocity (which will be set equal to one).
Disorder enters as a random abelian $U(1)$ vector potential $\sigb\cdot \vex{A} = \sigma_1 \, A^1 + \sigma_2 \, A^2$ 
with disorder strength $\lambda_A$; the overline $\overline{\cdots}$ denotes an average over disorder configurations.  
The delta function $\delta_\xi^\pup{2}(\vex{r} - \vex{r'})$ is smeared out by a correlation length $\xi$ 
in the numerics described below. 

\paragraph{Two Dirac nodes, U(1) $\oplus$ SU(2) vector potential dirt (AIII, $\nu=2$)\label{sec:nu=2}}
The winding number $\nu=2$ AIII model can be realized by adding a second Dirac cone (addressed by the color space $\tau_3 = \pm 1$). 
There is not only the Abelian U$(1)$ vector potential $\vex{A}_{0}$ with strength $\lambda_A$, but also a non-abelian $\text{SU}(2)$ vector potential $\vex{A}_{i}$ ($i \in \{1,2,3\}$) with strength $\lambda$:
\begin{align}\label{AIII-2}
\begin{gathered}
	H_{\mathsf{AIII}}^{\pup{2}}
	\equiv
	\sigb
	\cdot
	\left[
		-i \, \vv \, \Nabla
		+
		\vex{A}_{0}(\vex{r})
		+
		\vex{A}_{i}(\vex{r})
		\,
		{\tauh}_i
	\right],
\\
	\overline{A^a_0(\vex{r}) \, A^{b}_0(\vex{r'})} 
	=
	\lambda_A \, \delta^{a b} \, \delta_\xi^\pup{2}(\vex{r} - \vex{r'}),
\qquad
	\overline{A^a_i(\vex{r}) \, A^{b}_j(\vex{r'})} 
	=
	\lambda \, \delta^{a b} \, \delta_{i j} \, \delta_\xi^\pup{2}(\vex{r} - \vex{r'}).
\end{gathered}
\end{align}
In the limiting case of $\lambda_A=0$, the full $\text{SU}(2)$ spin symmetry as well as particle hole symmetry are restored. This puts the model into class CI, identical to the $\nu=2$ CI Hamiltonian in Eq.~\eqref{CI-2}.


\subsection{Multifractal analysis}

For the class AIII dirty Dirac theories described above, at zero energy (the surface quasiparticle Dirac point) 
the WZNW theory [Eq.~(\ref{WZNW}) with $\omega = 0$] predicts exact parabolicity with curvature 
[Eqs.~(\ref{Delta(q)Def}), (\ref{MFC_WZNW}), and Table~\ref{tab:WZW}]
\begin{align}
	\theta_\nu = \frac{|\nu|-1}{\nu^2}+ \frac{\lambda_A}{\pi},
\end{align}
depending on winding number $\nu$ and abelian disorder strength $\lambda_A$. 
With momentum-space exact diagonalization in Ref.~\cite{Chou2014}, 
the multifractal statistics of the low-energy states for the $\nu=1$ Hamiltonian \eqref{AIII-1} 
were shown to match this expression.

\begin{figure*}[t]
	\noindent \begin{centering}
		\includegraphics[width=\textwidth]{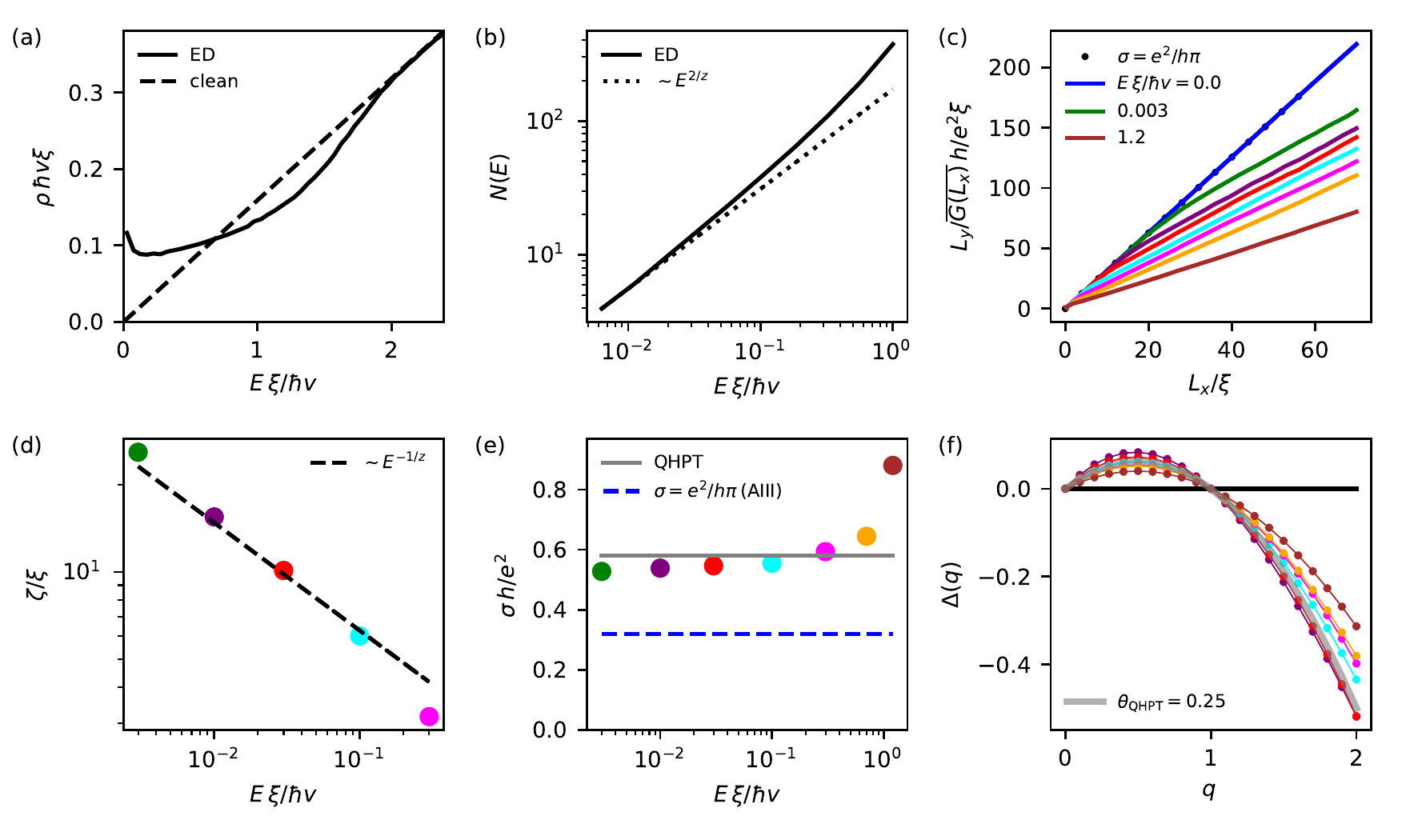}
		\par\end{centering}
	\caption{\label{fig:nu1}
		Numerical Landauer conductance and multifractal analysis from Ref.~\cite{Sbierski2020} 
		for the winding number $\nu=1$ AIII Hamiltonian, defined by Eq.~(\ref{AIII-1}). 
		The random abelian vector potential strength is 
		$\sqrt{\lambda_A} \equiv W = 2.3$.
		(a) The DOS $\rho(E)$ versus energy $E$, as calculated from momentum-space exact diagonalization (ED), 
		is most strongly affected by disorder around the Dirac point ($E = 0$). 
		(b) The integrated DOS $N(E) = \int_0^E d \e \, \rho(\e)$ is plotted versus energy. 
		The predicted scaling form implied by Eq.~(\ref{DoS_nu}) is governed by the disorder-dependent 
		dynamical critical exponent $z=1+W^{2}/\pi$. 
		(c) Quantum transport results for the resistance normalized to system width. 
		The energies are from top to bottom 
		$E\xi/\hbar \vv=0,0.003,0.01,0.03,0.1,0.3,0.7,1.2$.
		(d) The \emph{crossover correlation scale} from the transport calculation scales as
		$\zeta(E) \sim E^{-1/z}$. This scale (not to be confused with the fixed disorder
		correlation length $\xi$) governs the crossover at energy $E$ between WZNW and class A
		IQHPT criticalities at smaller and larger length scales, respectively. 
		(e) Conductivities extracted from the slope of the curves in panel (c), compared to the established
		value of the class A IQHPT critical conductivity (see Table~\ref{tab:WZW}). 
		(f) Anomalous part of the multifractal spectrum $\Delta(q)$ extracted from box-size scaling
		of ED eigenstates for box sizes beyond the crossover correlation length $\zeta(E)$,
		as extracted in (d). The data correspond to 
		$E\xi/\hbar \vv=0.01,0.03,0.1,0.3,0.7,1.2$
		(bottom to top).}
\end{figure*}

Ref.~\cite{Sbierski2020} further analyzes the conductance and the finite-energy multifractal properties. 
In Fig.~\ref{fig:nu1}(f) the multifractal spectrum of the $\nu=1$ model 
[Eq.~\eqref{AIII-1}] with linear size $L = 60 \xi$ is analyzed.  
Here $\xi$ denotes the common correlation length of the disorder potentials, 
which are taken to be Gaussian distributed and correlated \cite{Sbierski2020}. 
At finite energies, the anomalous multifractal spectrum $\Delta_q$ 
is compared to the parabolic approximation for the class A IQHPT 
with $\theta_{\mathsf{QHPT}}=0.25$ [Eqs.~(\ref{Delta(q)Def}), (\ref{MFC_WZNW}), and Table~\ref{tab:WZW}].
Great agreement is found over a wide energy range. In the high-energy tail $\e \sim 2\hbar \vv/\xi$, 
there are larger deviations. This can be explained since close to the energy cutoff the system seems 
untouched by the disorder and matches the clean DOS [see Fig.~\ref{fig:nu1}(a)]. 
Panel (b) in Fig.~\ref{fig:nu1} moreover confirms that the low-energy integrated DOS $N(\e)$ 
scales as expected from WZNW theory
[Eq.~(\ref{DoS_nu}) and Table~\ref{tab:WZW}].

\begin{figure}[t]
	\begin{centering}
		\includegraphics[width=\textwidth]{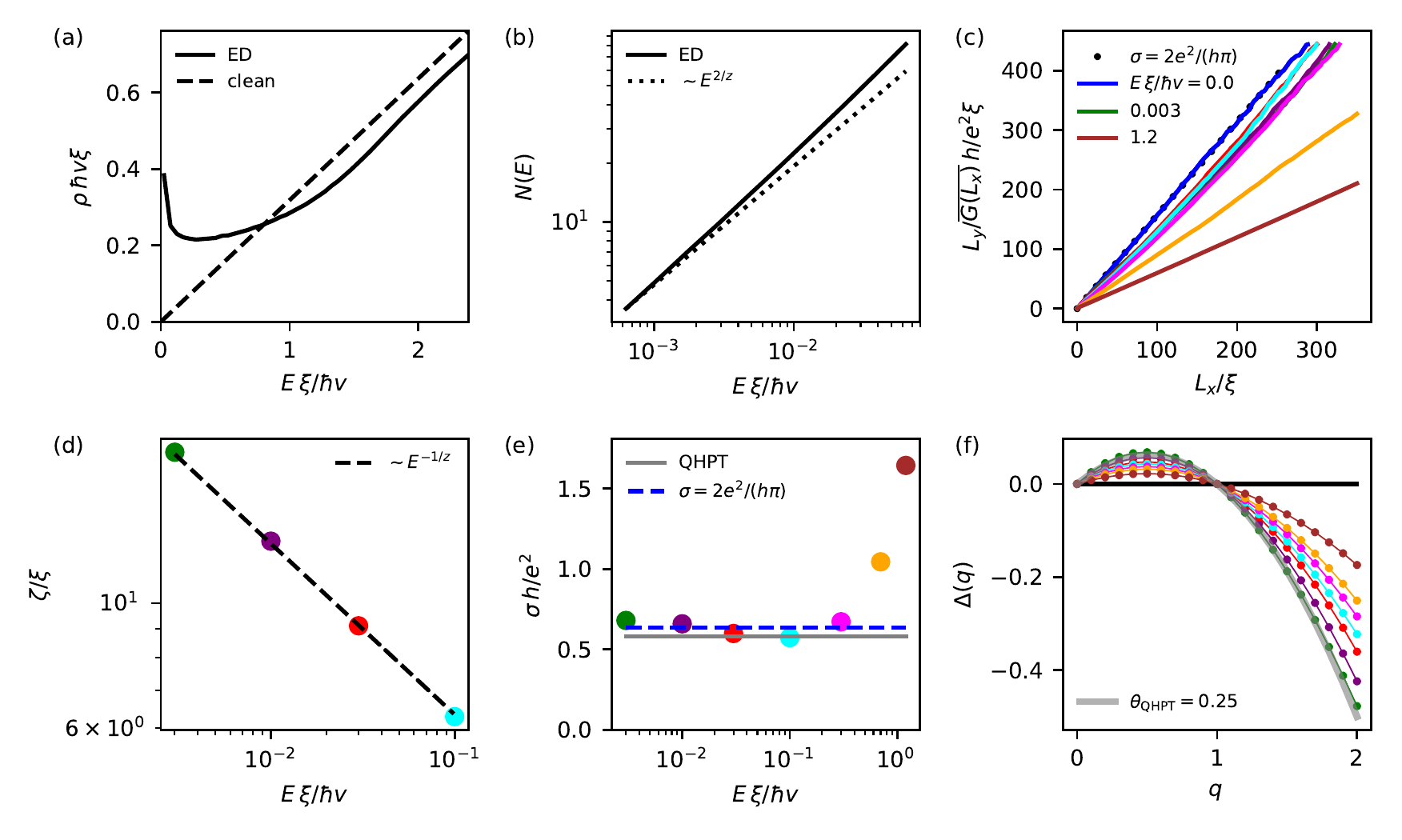}
	\end{centering}
	\caption{\label{fig:nu2}
		Numerical results from Ref.~\cite{Sbierski2020} for the 
		topological class AIII surface model with a two Dirac nodes $(\nu=2)$, defined by Eq.~(\ref{AIII-2}).
		The abelian and non-abelian vector potential disorder strengths are 
		$\sqrt{\lambda_A} \equiv W_{A} = 2.2$ 
		and 
		$\sqrt{\lambda} \equiv W_{N} = 1.5$, 
		respectively. 
		(a) The DOS as a function of energy, as calculated from ED. 
		(b) The integrated DOS $N(E) = \int_0^E d \varepsilon \, \rho(\varepsilon)$ plotted versus energy. 
		The predicted scaling form implied by Eq.~(\ref{DoS_nu}) is governed by the dynamical critical exponent $z = 7/4+W_{A}^{2}/\pi$,
		which depends only on the abelian disorder strength. 
		(c) Quantum transport results for the resistance normalized to system width. 
		The energies are from
		top to bottom $E\xi/\hbar \vv=0,0.003,0.01,0.03,0.1,0.3,0.7,1.2$. 
		(d)
		The crossover correlation length from the transport calculation scales as
		$\zeta(E) \sim E^{-1/z}$. 
		(e) Conductivities extracted
		from the $L_{x}\protect\geq200\xi$ slopes of the curves in panel
		(c), compared to the established value
		of the class A IQHPT critical conductivity (see Table~\ref{tab:WZW}).
		(f) Anomalous part of the multifractal spectrum $\Delta(q)$ extracted
		from box-size scaling of ED eigenstates for box sizes beyond the correlation
		length $\zeta(E)$ as extracted in (d). 
		The data correspond to $E\xi/\hbar \vv=0.003,0.01,0.03,0.1,0.3,0.7,1.2$
		(bottom to top).}
\end{figure}

The results for the \emph{$\nu=2$} surface theory \eqref{AIII-2} with the same linear size $L = 60 \xi$ are 
shown in Fig.~\ref{fig:nu2}. In panel (f) there is a comparison of finite-energy multifractal spectra to the 
class A IQHPT parabola with $\theta_{\mathsf{QHPT}}=0.25$. 
In the high-energy tail $\e\sim2\hbar \vv/\xi$ there are deviations for the same reason as in the $\nu=1$ case.

\noindent

\begin{figure}
	\includegraphics[width=\textwidth]{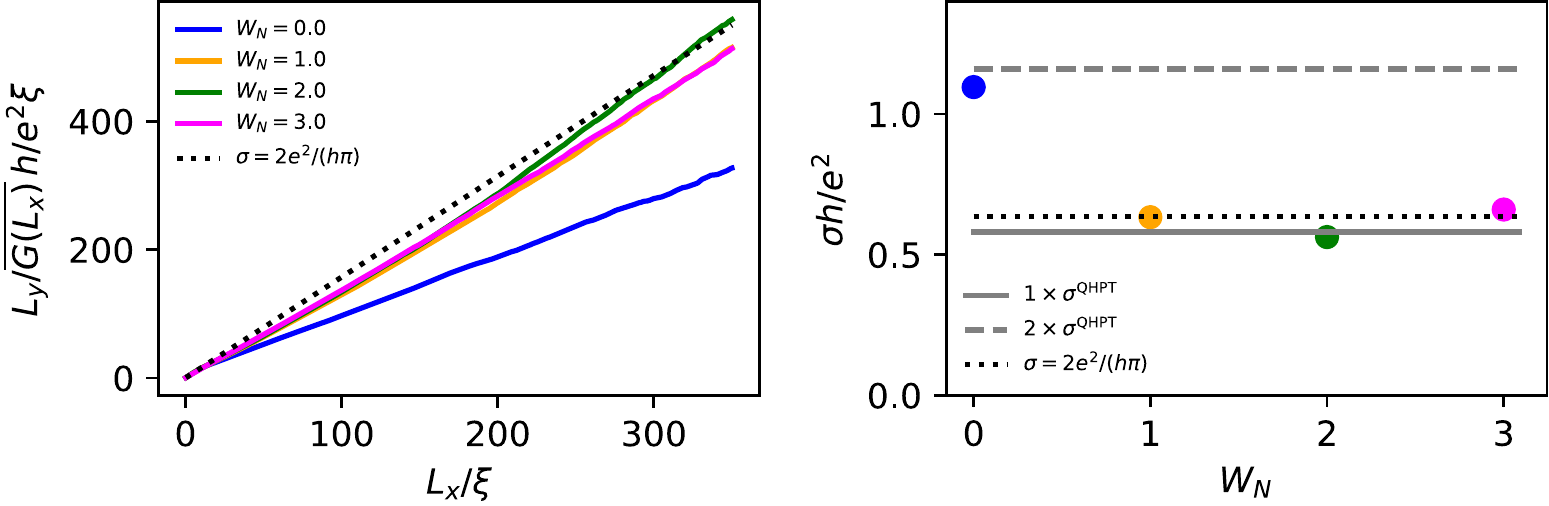}
	\caption{\label{fig:nu2-WNsweep}
		Numerical transport results from Ref.~\cite{Sbierski2020} 
		for the topological two-node class AIII Dirac model [$\nu = 2$, Eq.~(\ref{AIII-2})],
		with abelian and non-abelian vector potential disorder
		of strengths $W_{A}=2.1$ and increasing $W_{N}=0,1,2,3$ at energy
		$E=0.03\hbar \vv/\xi$. The left panel shows the bare resistance data,
		while the right panel depicts the bulk conductivities obtained from
		linear fits to the bare resistance data above $L_{x}=200\xi$.
		These plots establish the crossover of the two-node model from the
		finite-energy conductivity plateau equal to 
		$2\times\sigma^{xx}_{\mathrm{IQHPT}}$ in the absence of internode scattering,
		to a plateau with value 
		$1\times\sigma^{xx}_{\mathrm{IQHPT}}$ in its presence.
	}
\end{figure}


\subsection{Landauer conductance}

The transport calculations are performed by slicing the system in the $x$ direction 
and subsequently recasting the time-independent Schr\"odinger
equation $H_{\mathsf{AIII}}^{\pup{1}} \, \psi = \e \, \psi$ in terms of the transfer matrix, 
using the method of Ref.~\cite{Bardarson2007}. Clean, highly doped leads are attached to the system at 
$x = 0$ and $x = L_{x}$. The conductance $G$ is then computed from the transmission block $t$ 
of the scattering matrix $S$ between the leads.

The finite-size resistance normalized to the sample width $L_{y}/\overline{G}(L_x)$ is expected depend linearly on $L_x$ 
\begin{align}
	L_{y}/\overline{G}(L_x)=L_{y}R_{0}+\frac{1}{\sigma}L_{x}.
\end{align}
Gauge invariance and chiral symmetry force the contact resistance $R_0$ to zero for each configuration 
\cite{Ostrovsky2006,Schuessler2009}.
The data in Fig. \ref{fig:nu1}(c) for zero energy $E=0$ is consistent with the 
WZNW theory conductivity result 
$\sigma^{xx}_{\mathrm{AIII},\nu = 1} = e^{2}/ h \pi$, 
see Table~\ref{tab:WZW}. 
(Here the quoted conductance quantum $e^2/h$ is appropriate to charged electrons at the
surface of a chiral topological insulator in class AIII \cite{Hosur2010}. 
At the surface of a class AIII TSC, this should instead be replaced by the spin conductance
quantum $\hbar/8 \pi$ \cite{SRFL2008,Xie2015}.) 

The finite-energy crossover scale $\zeta(E)$ is defined as the length $L_{x}$ where $L_{y}/\overline{G}$
deviates by 5\% from the $E=0$ result. 
This is shown in Fig.~\ref{fig:nu1}(d). It follows the scaling $\zeta(E) \sim E^{-1/z}$, consistent with the 
$z$ determined from the DOS scaling. Physically, $\zeta(E)$ separates class AIII WZNW critical scaling
for shorter length scales from class A IQHPT scaling at larger ones; $\zeta(E) \rightarrow \infty$ 
as $E \rightarrow 0$. 

Finally the conductivity at finite energy is analyzed in Fig.~\ref{fig:nu1}(e). 
For $0<E\lesssim\hbar \vv/\xi$,
there is a plateau at 
$
\sigma 
\simeq 
0.55
(e^2/h)
$
in fair agreement with the value 
$
\sigma^{xx}_{\mathrm{IQHPT}}
=
0.58 \pm 0.02 \frac{e^{2}}{h}
$
obtained by Schweitzer and Marko\v s \cite{Schweitzer2005} via the
Kubo formula for a lattice model tuned to the class A IQHPT (Table~\ref{tab:WZW}). 
At larger
energies $E$, the conductivity at the accessible length scales increases
with energy.
This is expected for the semiclassical Drude conductivity,
which goes as $\sigma^{xx} \sim (e^2/h)(1/W^2)$, where $W^2$ is the disorder strength \cite{FosterAleiner2008}.
For these large
energies, the available length scales are insufficient to decide which
scenario, Anderson localization or IQHPT criticality, is realized at
the largest length scales.

Results for the $\nu = 2$ model are depicted in Fig.~\ref{fig:nu2}.
The largest scattering region of the \emph{$\nu=2$} sample is 
$(L_{x}=350\xi) \times (L_{y}=400\xi)$. The consistency check of the crossover length $\zeta$ in Fig.~\ref{fig:nu2}(c) works just as in the $\nu=1$ case. The dynamical critical exponent $z$ matches with the expected DOS scaling in fig.~\ref{fig:nu2}(b). The conductivity as function of energy shown Fig.~\ref{fig:nu2}(e) matches 
$\sigma^{xx}_{\mathrm{AIII},\nu = 2}$ for very low energies. 
At finite energies it slightly drops to $\sigma^{xx}_{\mathrm{IQHPT}}$. This drop cannot be resolved in the 
Kubo computations in Fig.~\ref{fig:aiii_kubo2} that we performed, discussed in the next subsection.

A complementary perspective on
the results in Fig.~\ref{fig:nu2} for the $\nu = 2$ surface theory is the following. 
This time, consider ramping up from zero the non-abelian disorder strength $W_N$. 
In Fig.~\ref{fig:nu2-WNsweep}, results for the conductivity are shown for 
$W_{N}=0,1,2,3$ at fixed abelian disorder strength $W_{A}=2.1$
and fixed energy $E=0.03\hbar \vv/\xi$. 
For $W_{N}=0$, the two nodes are decoupled
and the conductivity is close to $2\times\sigma^{xx}_{\mathrm{IQHPT}}$, 
as expected for two replicas of the single node case. 
For $W_{N}=1,2,3$ the nodes are coupled and the conductivity is close
to the value $1\times\sigma^{xx}_{\mathrm{IQHPT}}$.

Finally, we mention additional numerical evidence for the
IQHPT-stacking scenario obtained in Ref.~\cite{Sbierski2020}. 
First, the full Landauer conductance distribution 
was computed for both $\nu = 1,2$ class AIII Dirac models in Ref.~\cite{Sbierski2020}.
These were calculated for square samples of various sizes,
at several energies throughout the surface-state spectrum.  
The results for both $\nu = 1,2$ were found to be consistent
with the known distribution for the class A IQHPT \cite{JovWang1998}.
Second, the results for the ``anomalous'' (WZNW) Dirac
models defined by Eqs.~(\ref{AIII-1}) and (\ref{AIII-2})
were benchmarked against identical calculations
for nontopological class A and class AIII (Gade) 2D Dirac models.
The Gade model \cite{Guruswamy2000,GadeWegner1991,Gade1993}
arises as the continuum description of a 2D bipartite lattice model with
pure intersublattice hopping. 
These nontopological models were shown to exhibit clear signs of Anderson localization
at finite energy, as expected. Finally, we mention that multifractal
spectra were computed for surface states of a bulk 3D lattice model
for a class AIII TSC in Ref.~\cite{Sbierski2020}. Results
are consistent with those presented above. 
We do not reproduce these additional results here.


\subsection{Kubo conductivity}
\label{AIII-Kubo}

\begin{figure}[b!]
	\includegraphics[width=\textwidth]{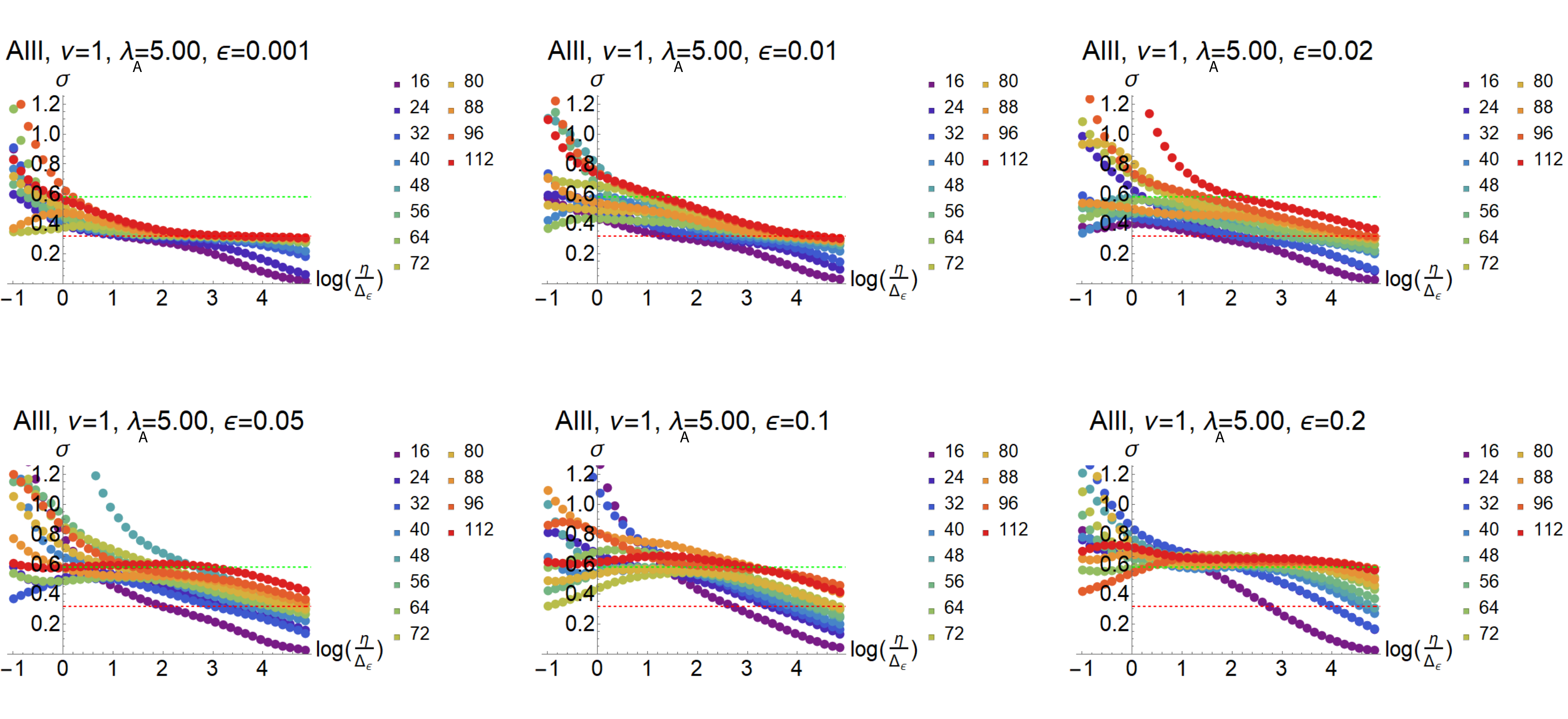}
	\caption{Numerical Kubo conductivity $\sk^{x x}$ computed with Eq.~\eqref{eq:num_kubo} for the 
	strongly disordered AIII $\nu=1$ model [Eq.~(\ref{AIII-1})], as function of the broadening $\eta$. 
	The disorder strength is $\lambda_A = 5$.
	A logarithmic scale for $\eta$ in units of the local level spacing $\Delta_\e$ is chosen. 
	There is convergence to a plateau of $\sk^{x x}$ as function $\eta$ as the linear system size $N$ increases. 
	At small energies $\e\ll\Lambda$, $\sk^{x x}$ tends to the WZNW value associated to the zero-energy state
	of the Dirac theory (red dashed). For finite energies $\e \lesssim \Lambda$, we find a value of $\sk^{x x}$ 
	compatible with the universal IQHPT result $\sigma^{xx}_{\mathrm{IQHPT}}$ (green dashed). States at 
	$\e\approx \Lambda$ are not affected much by the disorder and therefore do not show universal conductance values. 
	This confirms the Landauer computation in Fig.~\ref{fig:nu1}(e).}
	\label{fig:aiii_kubo1}
\end{figure}

\begin{figure}[b!]
	\includegraphics[width=\textwidth]{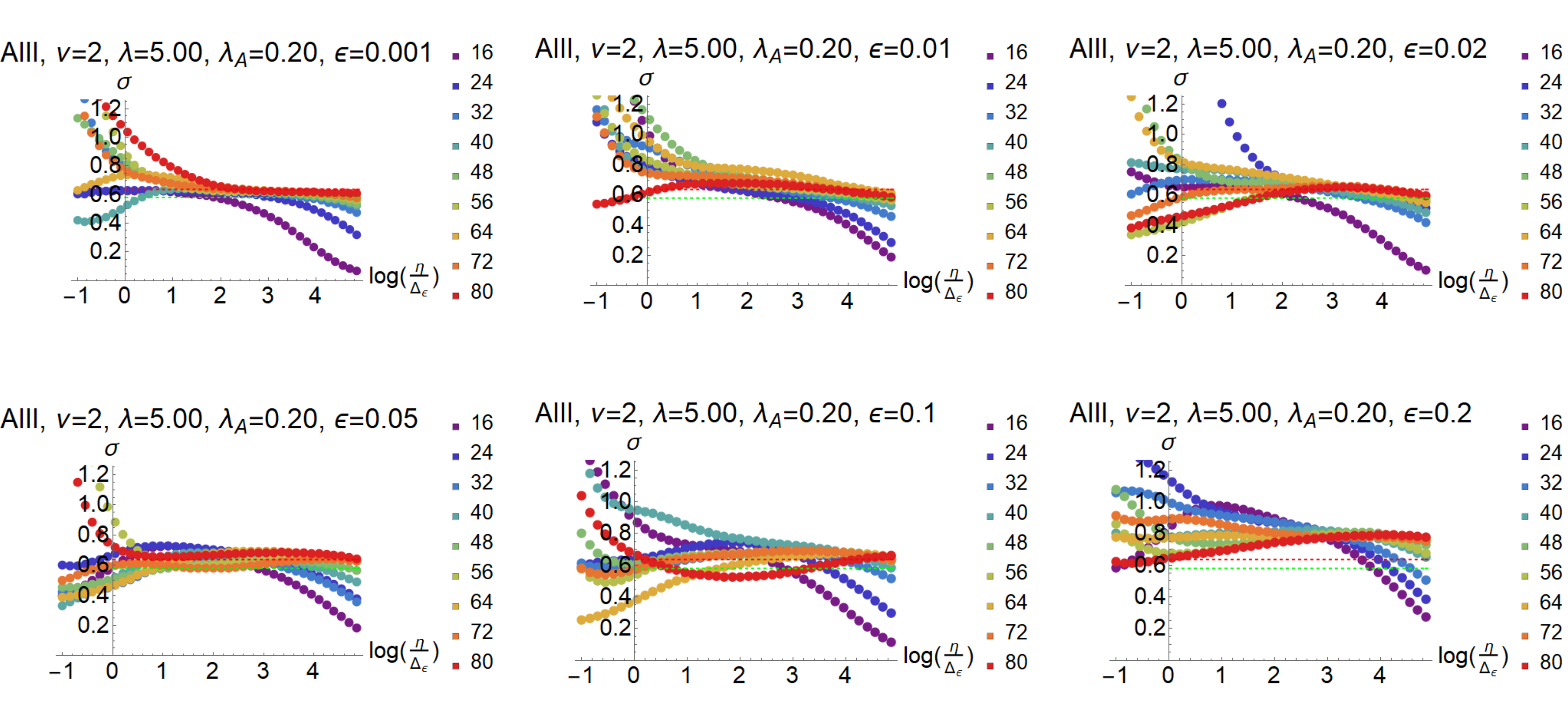}
	\caption{Same as Fig.~\ref{fig:aiii_kubo1} for the AIII $\nu=2$ model [Eq.~(\ref{AIII-2})]. 
	The nonabelian disorder strength is $\lambda = 5$, while the abelian strength is $\lambda_A = 0.2$. 
	The WZNW value (red dashed) and the  universal IQHPT result (green dashed) are very close to each other. 
	The convergence of $\sk^{x x}$ is not as clear as in Fig.~\ref{fig:aiii_kubo1}. Since 
	the model incorporates the 2D color space [Eq.~\eqref{AIII-2}], only systems with $N\leq 80$ are numerically 
	accessible. The data does not show signs of the conventionally expected Anderson localization 
	$\sk^{x x}\rightarrow 0$ at finite energies for this even-winding number class AIII system.
	In contrast to the more precise Landauer computation presented in Figs.~\ref{fig:nu2}(e) and \ref{fig:nu2-WNsweep}, 
	we cannot distinguish the zero- and finite-energy behavior. States at $\e\approx \Lambda$ are not 
	affected much by the disorder and therefore do not show universal conductance values.}
	\label{fig:aiii_kubo2}
\end{figure}

In this subsection, we present new results testing the AIII WZNW~$\rightarrow$~A IQHPT stacking conjecture,
this time computing the dc surface conductivity via the Kubo formula.
In natural units $e=1, \hbar=1$, the Kubo formula relates the dc conductivity $\sk^{a b}$ to the 
current-current response function
$\mathcal{K}^{a b}(\e)$:
\begin{align}
	\sk^{a b}
	&= 
	\dfrac{1}{4\pi}\int_{-\infty}^{\infty}
	\left[-\dfrac{df(\e)}{d\e}\right]
	\mathcal{K}^{a b}(\e), 
\\
	\mathcal{K}^{a b}(\e)
	&\equiv 
	\dfrac{1}{L^2}
	\int_{{\bf r}, {\bf r'}} 
	\Tr 
	\left[
		\hat{\sigma}^a
			\,
		\hat{\rho}(\e;{\bf r}, {\bf r'})
			\,
		\hat{\sigma}^b
			\,
		\hat{\rho}(\e;{\bf r'}, {\bf r})
		\right].
\end{align}
The finite size conductivity $\sk^{a b}$ must be evaluated with spectral densities 
$\hat{\rho}(\e;{\bf r}, {\bf r'})$ broadened by a finite $\eta$ of the order of the level spacing 
around energy $\e$:
\begin{align}
	\hat{\rho}(\e;{\bf r}, {\bf r'}) 
	&= 
	i\left[\hat{G}_R(\e;{\bf r}, {\bf r'}) - \hat{G}_A(\e;{\bf r}, {\bf r'})\right] 
	= 
	2\pi
	\sum_l 
	\left[
		\dfrac{\eta/\pi}{(\e-\e_l)^2+\eta^2}
	\right]
	\psi_l({\bf r}) \, \psi_l^\dagger({\bf r'}),
\end{align}
where $\psi_l(\vex{r})$ is an exact eigenstate. 
We employ 
\begin{align}
	\mathcal{K}^{a b}(\e)
	&=\left(\dfrac{2\pi}{L}\right)^2\sum_{l,m}\left[\dfrac{\eta/\pi}{(\e-\e_l)^2+\eta^2}\right]\left[\dfrac{\eta/\pi}{(\e-\e_m)^2+\eta^2}\right]\langle l |\hat{\sigma}^a|m\rangle \langle m |\hat{\sigma}^b|l\rangle \label{eq:num_kubo}
\end{align}
to compute the Kubo conductivity with eigenenergies $\e_l$ and states $|l\rangle$ from exact diagonalization. The result should be virtually independent of the broadening $\eta$ chosen around the local level spacing $\Delta_\e$. 
Calculations are performed for the momentum-space version of the continuum Hamiltonians
in Eqs.~(\ref{AIII-1}) and (\ref{AIII-2}), with quantized momenta corresponding to a finite-size torus 
and an ultraviolet energy cutoff $\Lambda$. 

We computed the Kubo conductivity $\sk^{x x}$ via Eq.~\eqref{eq:num_kubo} 
for the strongly disordered AIII \emph{$\nu=1$} model as a function of the level broadening $\eta$. 
A logarithmic scale for $\eta$ in units of the local level spacing $\Delta_\e$ is chosen 
(the critical DOS $\nu(\e)\propto\e^{\alpha}$ is responsible for the dependence $\Delta_\e\propto\e^{-\alpha}$). 
For large enough systems, $\sk^{x x}$ as function $\eta$ should depend only weakly on $\eta$. 
We use the tendency of $\sk^{x x}$ to converge towards a plateau value as a measure of finite size effects. 
The results are shown in Fig.~\ref{fig:aiii_kubo1}. At small energies $\e\ll\Lambda$, $\sk^{x x}$ tends towards 
the WZNW value $\sigma^{xx}_{\mathrm{AIII},\nu = 1} = 1/\pi$ (Table~\ref{tab:WZW}). 
For finite energies $\e \lesssim \Lambda$, we find a value of $\sk^{x x}$ compatible with the 
universal IQHPT result $\sigma^{xx}_{\mathrm{IQHPT}} \approx 0.58$ (green dashed). 
States at high energies $\e \approx \Lambda$ are only weakly affected by the disorder and 
therefore do not show universal conductance values. We are able to further confirm the Landauer 
computation in Ref.~\cite{Sbierski2020}, see Fig.~\ref{fig:nu1}(e).

For the AIII \emph{$\nu=2$} model, we show the results in Fig.~\ref{fig:aiii_kubo2}. The WZNW value 
$\sigma^{xx}_{\mathrm{AIII},\nu = 2} = 2/\pi$ (red dashed) and the IQHPT result $\sigma^{xx}_{\mathrm{IQHPT}} \approx 0.58$ (green dashed) are numerically close to each other. The convergence of $\sk^{x x}$ is not as clear as for the $\nu = 1$ case, Fig.~\ref{fig:aiii_kubo1}. Since we need an additional color space to realize this model [Eq.~\eqref{AIII-2}], only systems with $N\leq 80$ are numerically accessible. The data does not show signs of the conventionally expected Anderson localization $\sk^{x x} \rightarrow 0$ in the thermodynamic limit $N\rightarrow\infty$ at finite energies for $\nu = 2$ (i.e., even winding numbers). In contrast to the more precise Landauer computation performed in Ref.~\cite{Sbierski2020}, presented in Figs.~\ref{fig:nu2}(e) and \ref{fig:nu2-WNsweep}, we cannot distinguish the zero- and finite-energy behavior. Finite size effects estimated by the fluctuation of $\sk^{x x}(\eta)$ are larger than the numerical difference of the expected conductivities $\sigma^{xx}_{\mathrm{IQHPT}} \sim \sigma^{xx}_{\mathrm{AIII},\nu = 2}$. States at $\e \approx \Lambda$ are not affected much by the disorder and therefore do not show universal conductance values.


\begin{figure}[b!]
	\centering
	\includegraphics[width=\textwidth]{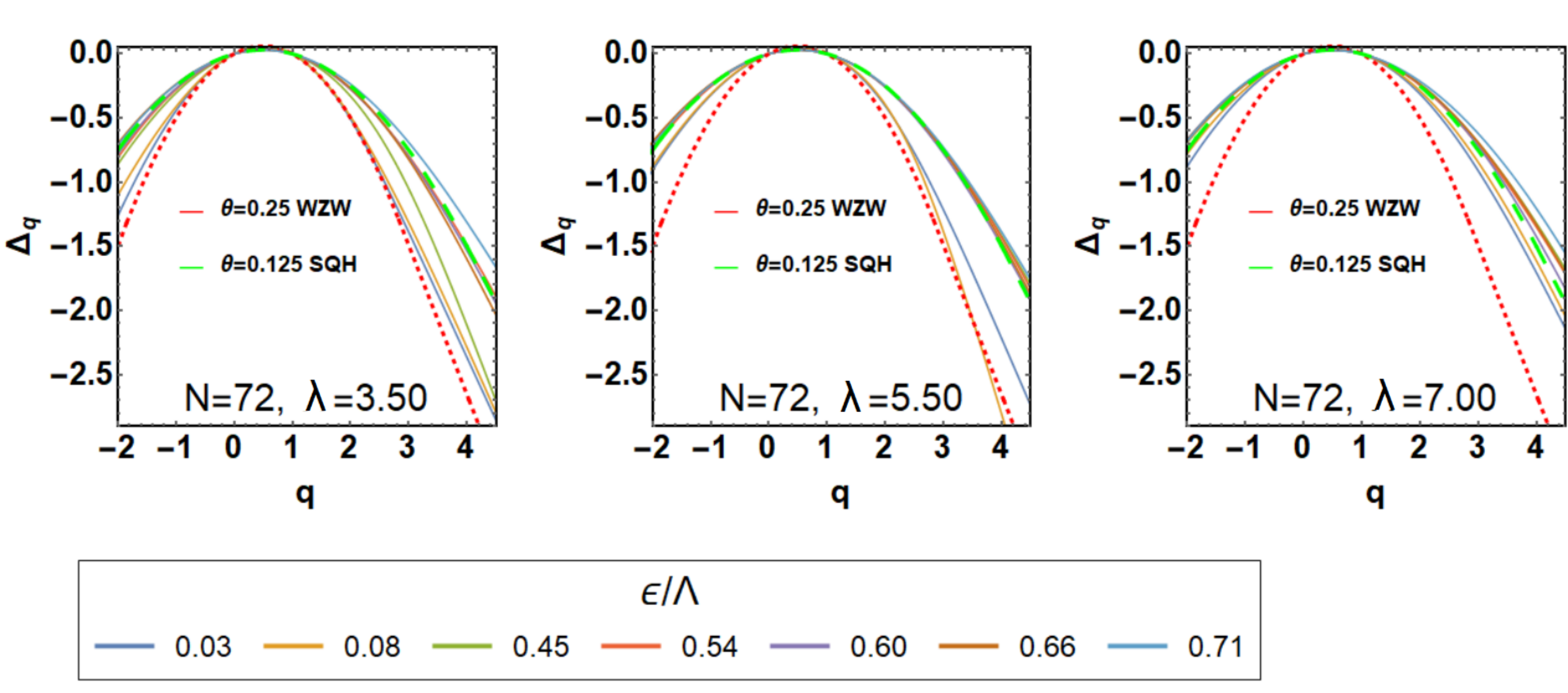}
	\caption{Multifractality in the CI $\nu=2$ model, 
	for the largest system size $N=72$, 
	at 
	weak $\lambda=3.5$, 
	intermediate $\lambda=5.5$, 
	and 
	strong disorder $\lambda=7.0$. 
	The system is a $(2 N + 1) \times (2N + 1)$ grid in momentum space; $\Lambda$
	is the ultraviolet energy cutoff for the clean Dirac spectrum.
	Near-zero energy states and finite-energy states are compared to the 
	class CI-WZNW (red dashed) and class C-SQHPT parabolic spectra (green dashed). 
	With increasing disorder, there are fewer and fewer states that match the CI-WZNW prediction,
	and the crossover scale moves towards zero energy. 
	Fig.~\ref{fig:multi_ci} exhibits a finite-size analysis of $\Delta_q$ for $q=2,3$.}
	\label{fig:parabola_ci}
\end{figure}

\begin{figure}
	\centering
	\includegraphics[width=0.85\textwidth]{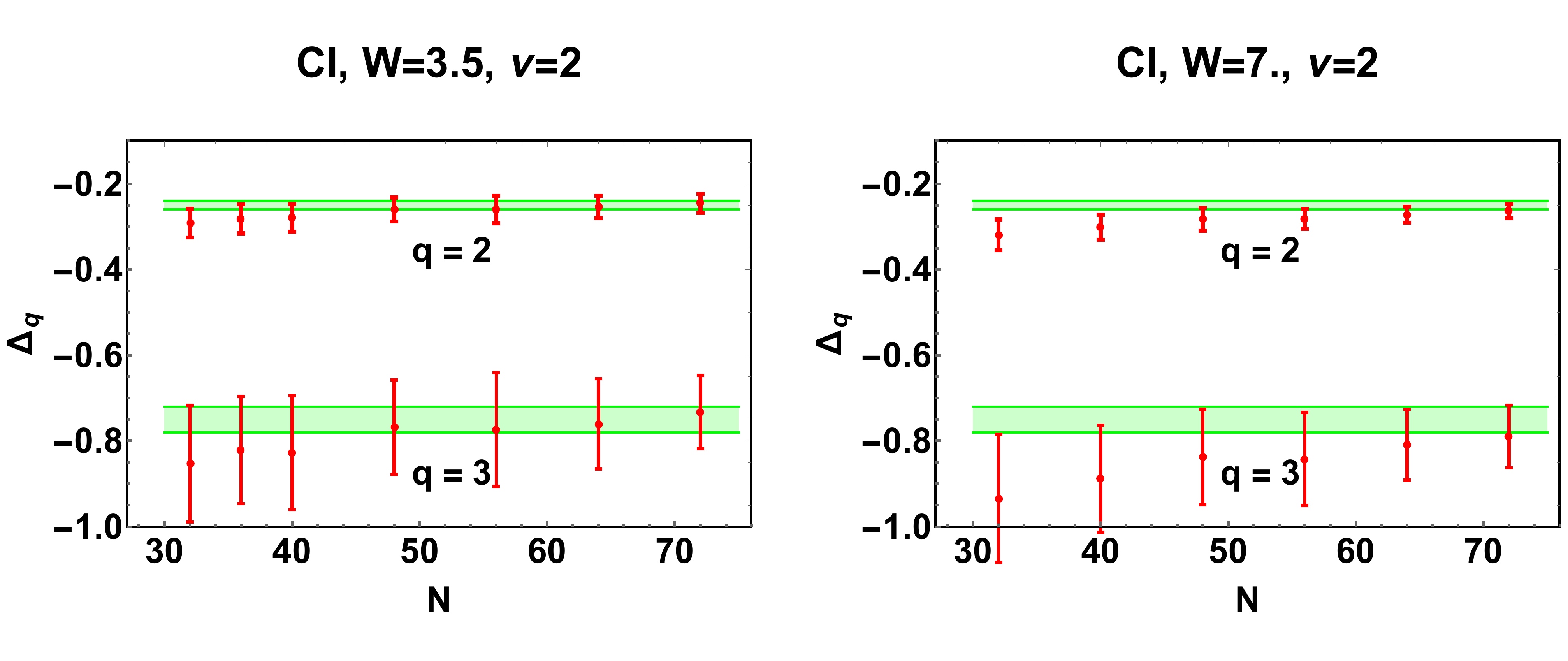}
	\caption{Finite energy $\Delta_q$ for $q=2,3$ in the class CI $\nu=2$ model as function of system size 
	$N=32,\ldots 72$. The green lines are exact analytical predictions 
	$\Delta_2= - {1}/{4}$, 
	$\Delta_3= - {3}/{4}$ for the class C SQHPT (see Table~\ref{tab:WZW}). 
	The red points are average values for $\Delta_q$ in the energy range 
	$0.2\lesssim \e/\Lambda \lesssim 1$; $\Lambda$ is the ultraviolet cutoff for the clean Dirac spectrum. 
	Error bars indicate the variance. When increasing $N$, 
	the $\Delta_q$ converge and fluctuations diminish for both intermediate and strong disorder $W\equiv\lambda$. }
	\label{fig:multi_ci}
\end{figure}

\begin{figure}
	\centering
	\includegraphics[width=0.85\textwidth]{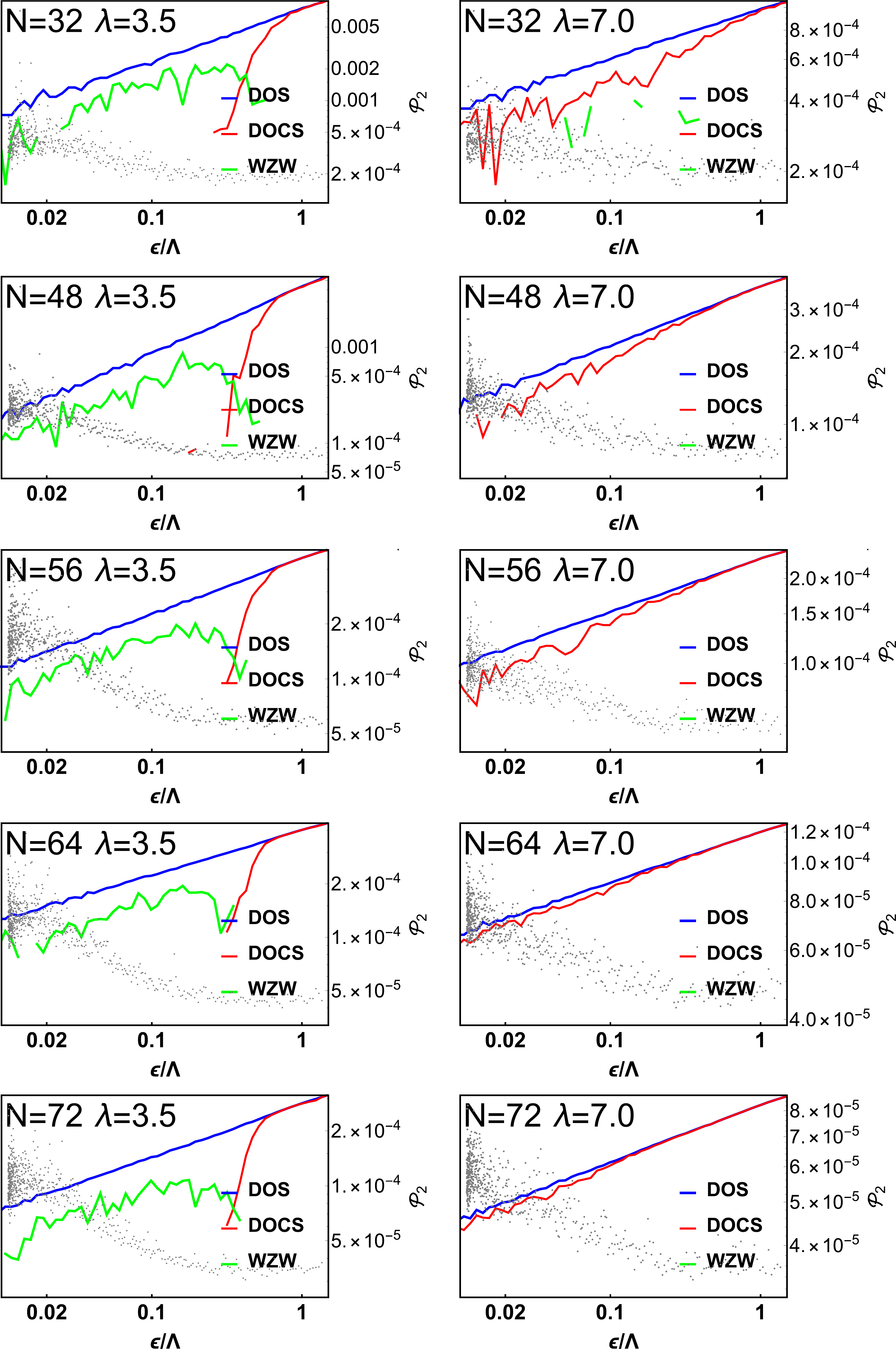}
	\caption{The density of critical states (DOCS) versus the total density of states (DOS) for the class CI model. 
	A state is termed critical when it matches the expected multifractal spectrum $\tau_q$ 
	for the class C-SQHPT within 4\% for at least 75\% of the $0<q<q_c$, where $q_c = 4$. 
	At zero energy we expect class CI WZNW-criticality and at finite energies SQHPT-criticality, 
	see Fig.~\ref{fig:multi_ci}. With increasing system size $N$ or disorder strength $\lambda$, 
	the amount of WZNW-critical states decreases in favor of class C-SQHPT critical finite-energy states. 
	(The green curve labeled ``WZW'' denotes the density of critical class CI-WZNW states). 
	Superimposed in light gray is the inverse-participation ratio $\mathcal{P}_2$,
	which shows that states away from zero energy are \emph{less} rarified, 
	as predicted by the stacking conjecture.}
	\label{fig:docs_ci}
\end{figure}

\begin{figure}[t!]
	\centering
	\includegraphics[width=\linewidth]{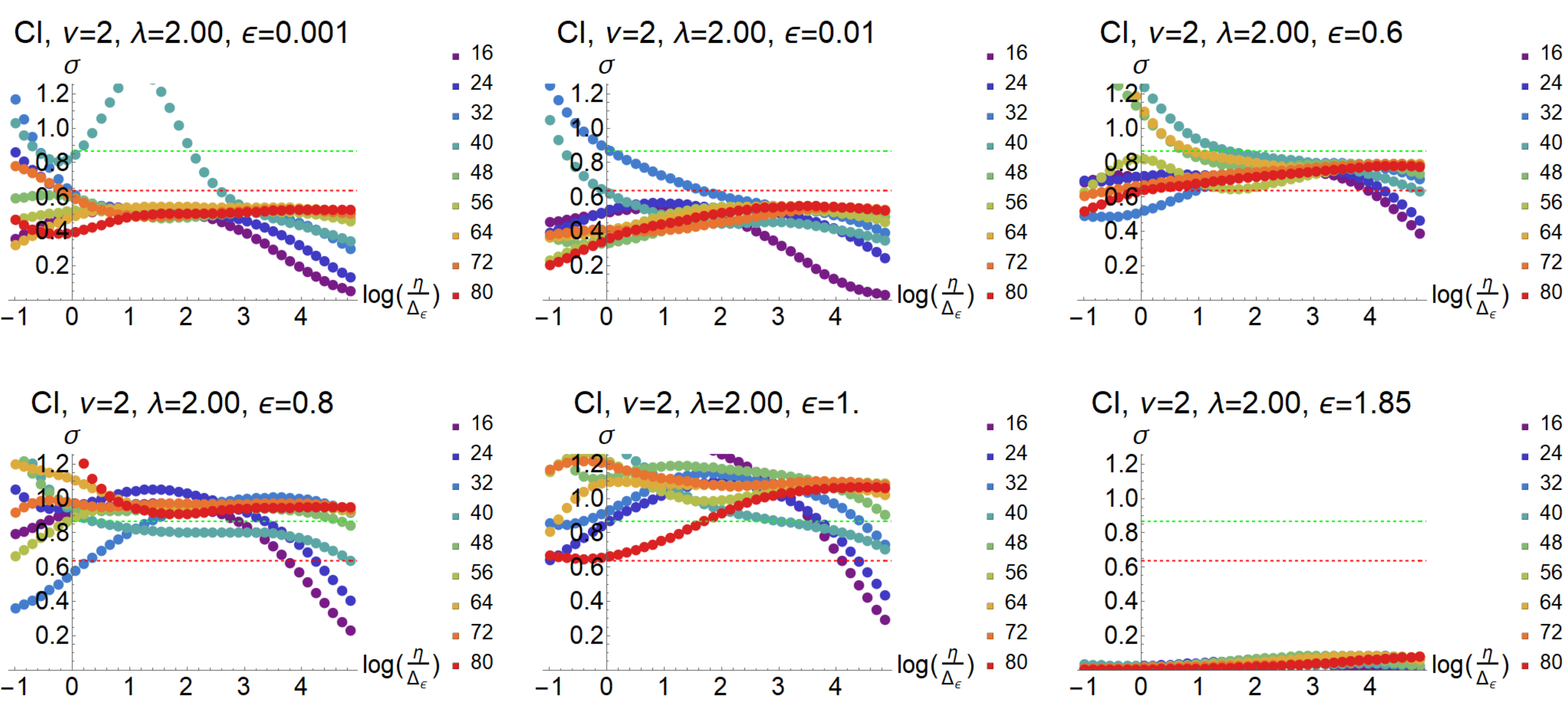}
	\includegraphics[width=\linewidth]{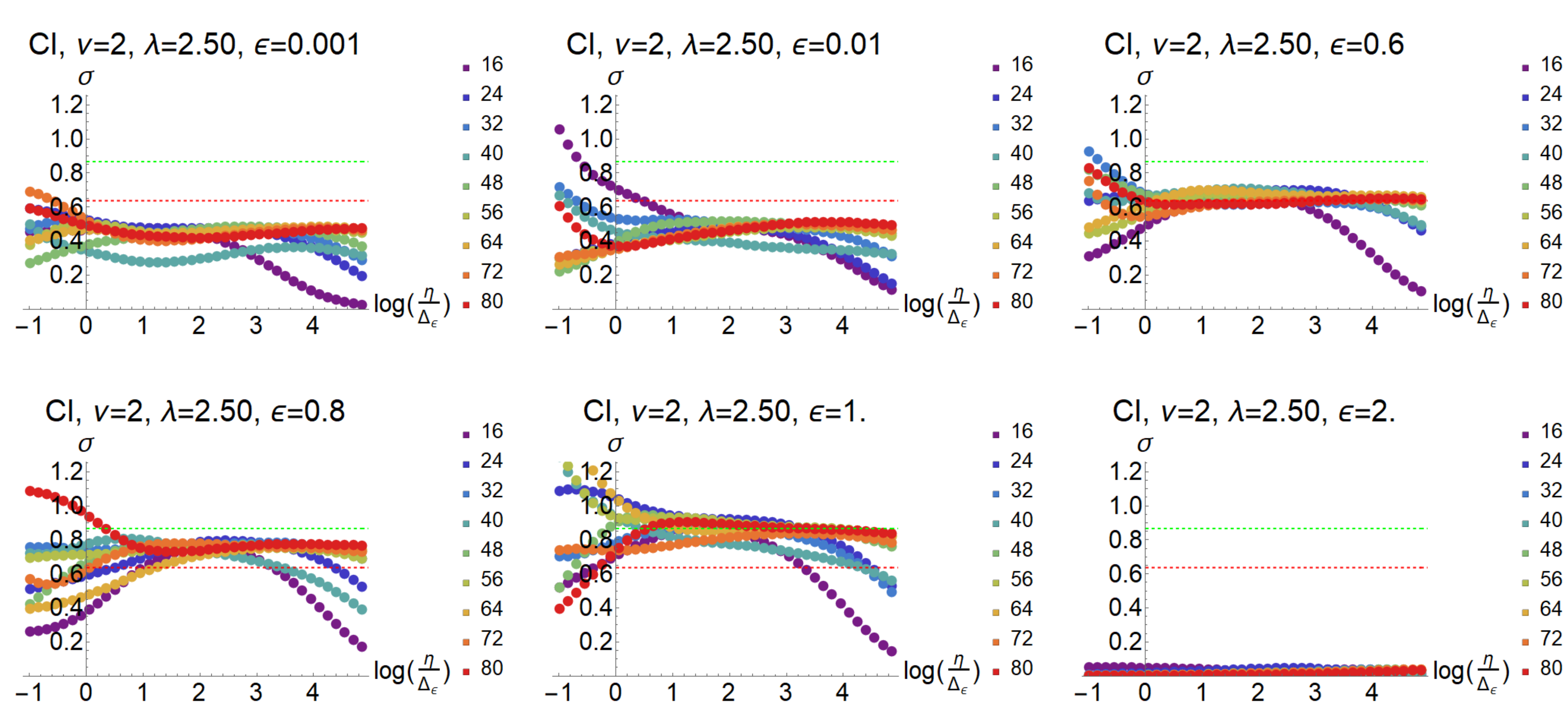}
	\caption{Numerical Kubo conductivity $\sk^{x x}$ for the class CI model, 
	computed with Eq.~\eqref{eq:num_kubo}, for moderate disorder as function of the level broadening $\eta$. 
	A logarithmic scale for 
	$\eta$ in units of the local level spacing $\Delta_\e$ is chosen. 
	Numerical results are compared to the 
	exact zero-energy WZNW result 
	$\sigma^{xx}_{\mathrm{CI},\nu = 2} = 2/\pi$ (red dashed) 
	and to the exact average value for the class C SQHPT 
	$\sigma^{xx}_{\mathrm{SQHPT}}={\sqrt{3}}/{2}$ (green dashed),
	see Table~\ref{tab:WZW}. 
	For most energies 
	(except the smallest), there is convergence as a function of $\eta$ as $N$ increases, 
	and $\sk^{x x}$ does not depend on $\eta$ significantly, i.e.\ shows a plateau. 
	For our purposes only the finite 
	energies are crucial, but at strong disorder $\lambda \gtrsim 3$ convergence also becomes poor there. 
	The increased finite-size effects near zero energy disable us from going to the strongly disordered regime, 
	where more of the spectrum is SQHPT-critical, according to the multifractal analysis 
	presented in Figs.~\ref{fig:parabola_ci}--\ref{fig:docs_ci}.
	Crucially, though, we only observe evidence for Anderson localization deep in the high-energy
	Lifshitz tail. This again contradicts the conventional picture that all finite-energy
	states localize in the orthogonal class AI (Secs.~\ref{sec:topclass} and \ref{sec:wzw}).
	}
	\label{fig:conductivity_ci}
\end{figure}

\section{Full SU(2) spin symmetry: class CI}
\label{sec:ci}

A class CI topological surface can be described by the following Dirac Hamiltonian:
\begin{align}\label{CI-2}
	H_{{\mathrm{CI}}} = \sigb\cdot\left[(-i \Nabla) + \vex{A}_i(\vex{r}) \, \tauh^i\right],
	\quad
	\overline{A^a_i(\vex{r}) \, A^{b}_j(\vex{r'})} 
	=
	\lambda \, \delta^{a b} \, \delta_{i j} \, \delta_\xi^\pup{2}(\vex{r} - \vex{r'}),
\end{align}
where $\sigb \equiv \sigh^1 \, \hat{x} + \sigh^2 \, \hat{y}$. The winding number $\nu = 2k$
is always even for class CI; disorder that preserves physical time-reversal and spin SU(2)
symmetry appears as the color gauge potential $\vex{A}_i$, where the $2 k \times 2 k$ 
color-space matrices $\{\tauh^i\}$ generate the color group Sp($2k$) \cite{Foster2014,Foster12,Schnyder2009}. 
For the minimal case $k = 1$, Eq.~(\ref{CI-2}) is identical to the $\nu = 2$ class AIII
Hamiltonian with vanishing abelian disorder [$\lambda_A = 0$ in Eq.~(\ref{AIII-2})].

In Ref.~\cite{Ghorashi2018}, extensive studies for several values of $\nu$ lead to the 
class CI WZNW~$\rightarrow$~C spin quantum Hall plateau transition (SQHPT) stacking conjecture.
Instead of employing Sp$(2 k)$ generators, for $k > 1$ a dispersion-modified version of the 
Hamiltonian in Eq.~(\ref{AIII-2}) was used to study class CI with higher winding numbers in 
\cite{Ghorashi2018}. 
In lieu of reproducing this data, here we perform new simulations for 
larger systems ($N_{max}=72$ vs.\ $N_{max}=46$), but focus on the minimal winding number $\nu=2$.
In this case we use Eq.~(\ref{CI-2}), where the color generators $\{\tauh^i\}$ are Pauli matrices.

\subsection{Multifractal analysis}

We numerically compute the multifractal spectrum $\Delta_q$  [Eqs.~(\ref{tau(q)Def}) and (\ref{Delta(q)Def})] 
using exact diagonalization of the continuum Dirac Hamiltonian in Eq.~(\ref{CI-2}) in momentum space. 
Results are shown for $\Delta_q$, computed for the largest available system size, 
at various energies in Fig.~\ref{fig:parabola_ci}. With increasing disorder strength $\lambda$, fewer and fewer 
states match the zero-energy class CI WZNW prediction [Eq.~(\ref{MFC_WZNW}) with $\theta = 1/4$], and
instead converge towards the parabolic approximation to the spin QHPT spectrum
with $\theta_{\mathsf{SQHPT}}=0.125$ (see Table~\ref{tab:WZW}). 
In Fig.~\ref{fig:multi_ci} we show $\Delta_q$ for $q = 2,3$ at finite energies $\e/\Lambda \gtrsim 0.2$. 
The red dots mark the average values over that part of the spectrum with the standard deviation given by the error bars. 
With increasing system size, the error bars shrink and there is convergence towards the $\theta_{\mathsf{SQHPT}}=1/8$ 
parabola. In contrast with the conventional reduction to the orthogonal Wigner--Dyson class AI at finite energy
(which would imply Anderson localization of all finite-energy states, see Secs.~\ref{sec:topclass} and \ref{sec:wzw}), 
we instead see evidence for ``stacked'' universal quantum criticality, consistent with the SQHPT. 

Following Ref.~\cite{Ghorashi2018}, in Fig.~\ref{fig:docs_ci} we statistically analyze stacking throughout
the energy spectrum. In each panel, we plot the total density of states (DOS), as well 
as the \emph{density of critical states} (DOCS). The latter is defined as follows. 
The DOCS is determined by the proportion of critical states at finite energy matching the
class C SQHPT $\theta_{\mathsf{SQHPT}}=1/8$ parabolic ansatz, within the tolerance criterion that 75\% of 
the $\tau_q$ with $q\in[0,q_c]$ have to match the parabolic $\tau_q^\theta$ up to $4\%$ accuracy.
Here $q_c(\theta) \equiv \sqrt{2/\theta}$ is the termination threshold \cite{Evers2008,Chamon1996,Ghorashi2018}.
In addition to the DOS and SQHPT DOCS, in Fig.~\ref{fig:docs_ci} we also exhibit the DOCS for
matching the zero-energy class CI WZNW prediction [Eq.~(\ref{MFC_WZNW}) with $\theta = 1/4$,
Table~\ref{tab:WZW}]. As indicated by the results in Fig.~\ref{fig:docs_ci}, 
more (less) of the spectrum matches the class C SQHPT (class CI WZNW) prediction as the 
system size $N$ or disorder strength $\lambda$ is increased. Since
the class C SQHPT states exhibit \emph{weaker} multifractality than the CI WZNW states,
this is strong evidence against Anderson localization.

\subsection{Kubo conductivity}

We compute the Kubo conductivity the same way as for the AIII surfaces in Sec.~\ref{AIII-Kubo}. 
In Fig.~\ref{fig:conductivity_ci}, we show the numerical Kubo results for selected energies across 
the spectrum. Near zero energy, we expect the exact WZNW result 
$\sigma^{xx}_{\mathrm{CI},\nu = 2} = 2/\pi$ (Table~\ref{tab:WZW}).
This result holds for \emph{every} disorder configuration, in the infinite system-size limit. 
For weak disorder $\lambda\lesssim 1$ we can observe this in the numerics. 
Finite-size effects seem to grow quickly with increasing disorder, 
which makes the most interesting regime $\lambda\gtrsim 3$, where SQHPT multifractality has spread over 
a wide range of the energy spectrum, difficult to reach numerically. 
This is the reason for the derivation from the exact WZNW result for $\lambda=2,2.5$ shown. 
At higher energies, the multifractal spectrum suggests class C SQHPT criticality. 
The average value of the conductivity is known to be 
$\sigma^{xx}_{\mathrm{SQHPT}}={\sqrt{3}}/{2}$ \cite{SQHPT-4_Cardy2000}. 

Although finite-size effects remain relevant as near zero energy, 
the results indicate convergence towards a finite conductivity, coincident with the Cardy value, at finite energies. 
This is again in contrast with the conventional expectation of Anderson localization
in the orthogonal class AI (Secs.~\ref{sec:topclass} and \ref{sec:wzw}).
On the other hand, for very high energies $\e\gtrsim \Lambda$ (in the Lifshitz tail), 
we observe localization in $\Delta_q$. This is consistent with the 
$\sk^{x x} \rightarrow 0$ here, as shown for energy $\e = 2 \Lambda$ in 
Fig.~\ref{fig:conductivity_ci}.


\begin{figure}[t!]
	\centering
	\includegraphics[width=\textwidth]{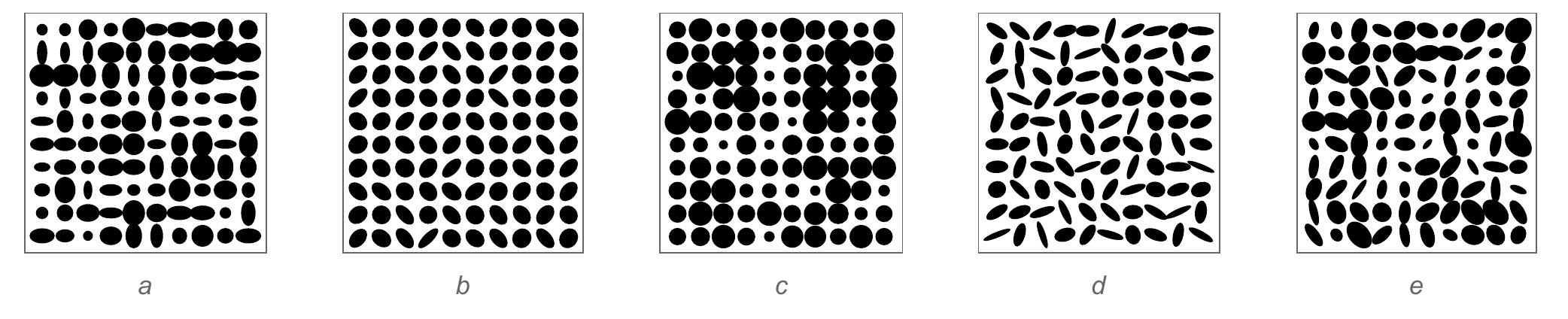}
	\caption{Visualization of quenched gravitational disorder (QGD): 
			the spatial components of the random metric tensor projected on a flat 2D space. 
			A 2$\times$2 tensor $v_{i j}$ can be visualized by the quadratic form $\{x : v_{ij} x_i x_j = r^2\}$ for some fixed $r>0$. 	
			We consider different classes of disorder (a)--(e) (see text):
			 (a) flattening/steepening + nematic disorder $v_{21} =v_{12} = 0$, 
			 (b) rotations + nematic disorder $\delta v_{11} = \delta v_{22} = 0$, 
			 (c) flattening/steepening + rotations $\delta v_{11} = \delta v_{22}$ and $\delta v_{12} = -\delta v_{21}$ (conformal spin $s=0$),
		     (d) full nematic disorder $\delta v_{11} = -\delta v_{22}$ and $\delta v_{12} = \delta v_{21}$ (conformal spin $s=-2,2$),
			 (e) generic disorder.
	}
	\label{fig:diii_metric}
\end{figure}

\section{Broken spin rotation symmetry: class DIII}
\label{sec:diii}

\subsection{Majorana surface fluids, quenched gravitational disorder, and possible relevance to high-$T_c$ cuprates \label{sec:QGD}}

The simplest bulk topological superconductor resides in class DIII,
with the minimal winding number $\nu = 1$; this would be a solid-state
analog of $^3$He-$B$ \cite{SRFL2008,Volovik,HeliumRev}, 
as has been proposed e.g.\ in 
Cu$_x$Bi$_2$Se$_3$ \cite{CuBiSe_pairing,Ando11,Chu13,Stroscio13,Zheng16,Yonezawa17,Tao18}
Nb$_x$Bi$_2$Se$_3$ \cite{Smylie16,Qiu17,Smylie17,Smylie20}, 
and 
$\beta$-PdBi$_2$ \cite{Kolapo18,Chien19}.
In this case, superconductivity and strong spin-orbit coupling imply
that neither charge nor spin transport is well-defined at the surface. 

The surface theory consists of a single massless Majorana cone. 
In contrast to the surface Hamiltonian in Eq.~(\ref{eq:surface}) 
(which applies for class DIII with winding numbers $|\nu| \geq 3$),
the only continuous symmetry available to gauge by disorder is Poincar\'e 
invariance, i.e.\ ``gravitational'' coupling to the stress tensor. 

Here we review the results of Ref.~\cite{Ghorashi2020}, in which 
the effects of ``quenched gravitational disorder'' (QGD) for a single
2D cone were studied. Time-reversal invariant perturbations, such as a charged
impurity, couple only to the spatial--spatial components of the stress--energy
tensor $T^{a b}$, with $a,b \in \{1,2\}$ \cite{Ghorashi2020}.
The most generic surface Hamiltonian takes the form
\begin{align}\label{DiracRVH}
H
=
-
\frac{1}{2}
\sum_{a,b = 1,2}
\int
d^2r
\,
v_{a b}(\vex{r})
\left(
\bar{\psi}
i
\sigh^a
\!
\stackrel{\leftrightarrow}{\parr_b}
\!
\psi
\right),
\end{align}
where the bidirectional derivative 
$A \!\! \stackrel{\leftrightarrow}{\vspace{-12pt}\parr} \!\! B = A \parr B - (\parr A) B$.
For the Majorana surface theory, $\bar{\psi} = \psi^\T \sigh^1$.
The \emph{four} velocity components are the isotropic Fermi velocity of the clean Majorana cone, 
perturbed by quenched random fluctuations:
$	\big\{	\voo(\vex{r}) \equiv 1 + \dvoo(\vex{r})$, 
$	\vtt(\vex{r}) \equiv 1 + \dvtt(\vex{r})$, 
$	\vot(\vex{r})$,		
$	\vto(\vex{r})	\big\}$.
In Ref.~\cite{Ghorashi2020}, five different variants of that model were considered. 
The variants are visualized in Fig.~\ref{fig:diii_metric},
\begin{enumerate}
	\item[(a)]{Independent $\{\dvoo,\dvtt\}$, $\vot=\vto = 0$.
		Local isotropic flattening or steepening of the Dirac cone and nematic squishing of the cone. 
	}
	\item[(b)]{Independent $\{\vot,\vto\}$, $\dvoo=\dvtt = 0$.
		Local pseudospin rotations (antisymmetric part ${\vot}_a=-{\vto}_a$) 
		and nematic squishing of the Dirac cone (symmetric part ${\vot}_s=+{\vto}_s$). 
	}
	\item[(c)]{Independent $\{\dvoo = \dvtt,\;\vot = - \vto\}$.
		Local isotropic flattening or steepening of the Dirac cone and pseudospin rotations. 
	}
	\item[(d)]{Independent $\{\dvoo = -\dvtt,\;\vot = \vto\}$.
		Local nematic squishing of the Dirac cone.
	}
	\item[(e)]{Independent $\{\dvoo,\dvtt,\vot,\vto\}$.
		The generic model.
	}
\end{enumerate}

Without further restrictions, the generic theory (e) is realized. 
For a fully isotropic bulk superfluid, it can be shown that electric potentials 
couple only through the isotropic flattening or steepening of the surface
Majorana cone, model (c) \cite{Ghorashi2020}. Crystal field effects will however
generically enable off-diagonal QGD (nonzero $\{\vot,\vto\}$).
Model (d) is another interesting special case, since pure nematic QGD couples
only to the holomorphic $T(z)$ and antiholomorphic $\bar{T}(\bar{z})$ stress tensor
components (using the language of 2D conformal field theory) \cite{Ghorashi2020}.

We also emphasize that QGD as in Eq.~(\ref{DiracRVH}) will generically be present
in \emph{any} 2D massless Dirac material. At zero energy (the Dirac point),
short-range correlated QGD is strongly irrelevant. This is why it is typically
ignored, compared to mass, scalar, or vector potential perturbations 
[as in Eq.~(\ref{eq:surface})]; short-ranged correlated disorder in the latter is
marginal at tree level. The very surprising finding in \cite{Ghorashi2020}, reviewed
below, is that while weak QGD is indeed irrelevant near zero energy, it appears
to induce quantum-critical stacking of states with weak, but universal multifractality,
similar to the class AIII and CI systems studied above. This occurs because
nonzero energy is a strongly relevant perturbation to the (2+0)-D Dirac-point theory.
For QGD, the latter can also be cast as a modified version of the WZNW model 
in Eq.~(\ref{WZNW}) with $\nu = 1$ and $\lambda_A = 0$; again $\omega \neq 0$ drives this model
away from the zero-energy WZNW fixed point (in this case, equivalent to free fermions),
towards some strong coupling regime. The numerical results presented below
indicate that this is another version of the critical stacking scenario. 

QGD might be important in the high-$T_c$ cuprate superconductors \cite{Ghorashi2020}. 
As reviewed in Sec.~\ref{sec:topclass}, the generic Bogoliubov--de Gennes Hamiltonian
for a 2D $d$-wave superconductor with non-magnetic disorder resides in the
non-topological version of class CI; in contrast to Eq.~(\ref{eq:surface}),
the low-energy Dirac theory for the 4 independent quasiparticle colors 
features mass, scalar, and vector potentials. This model is known 
to Anderson localize at all energies \cite{AltlandSimonsZirnbauer2002}. 
If however we assume that sufficiently long-wavelength disorder dominates, which 
does not scatter between the colors, then one obtains four independent copies
of the class AIII $\nu = 1$ WZNW Hamiltonian in Eq.~(\ref{AIII-1}) \cite{AltlandSimonsZirnbauer2002}.
Given the class AIII WZNW~$\rightarrow$~class A IQHPT stacking conjecture,
this would imply relatively strong multifractality (wave function rarification)
at finite energy, along with a strongly renormalized low-energy density of states.
This is not seen in experiment. However, \emph{nematic} QGD [as in models (b) and (d),
described above] in fact produces a phenomenology similar to that observed
in STM maps of the local density of states in BSCCO 
\cite{Ghorashi2020,Davis01,Davis02,Davis05-a,Davis05-b,Davis08,DavisReview}.
This includes plane-wave like states at low energy, with a linear-in-energy
density of states, but energy-independent, nanometer-scale critical
inhomogeneity at finite energies \cite{Ghorashi2020}. See also 
Fig.~\ref{fig:stacked} and Table~\ref{tab:WZW}.

\begin{figure}[b!]
	\centering
	\includegraphics[width=0.85\textwidth]{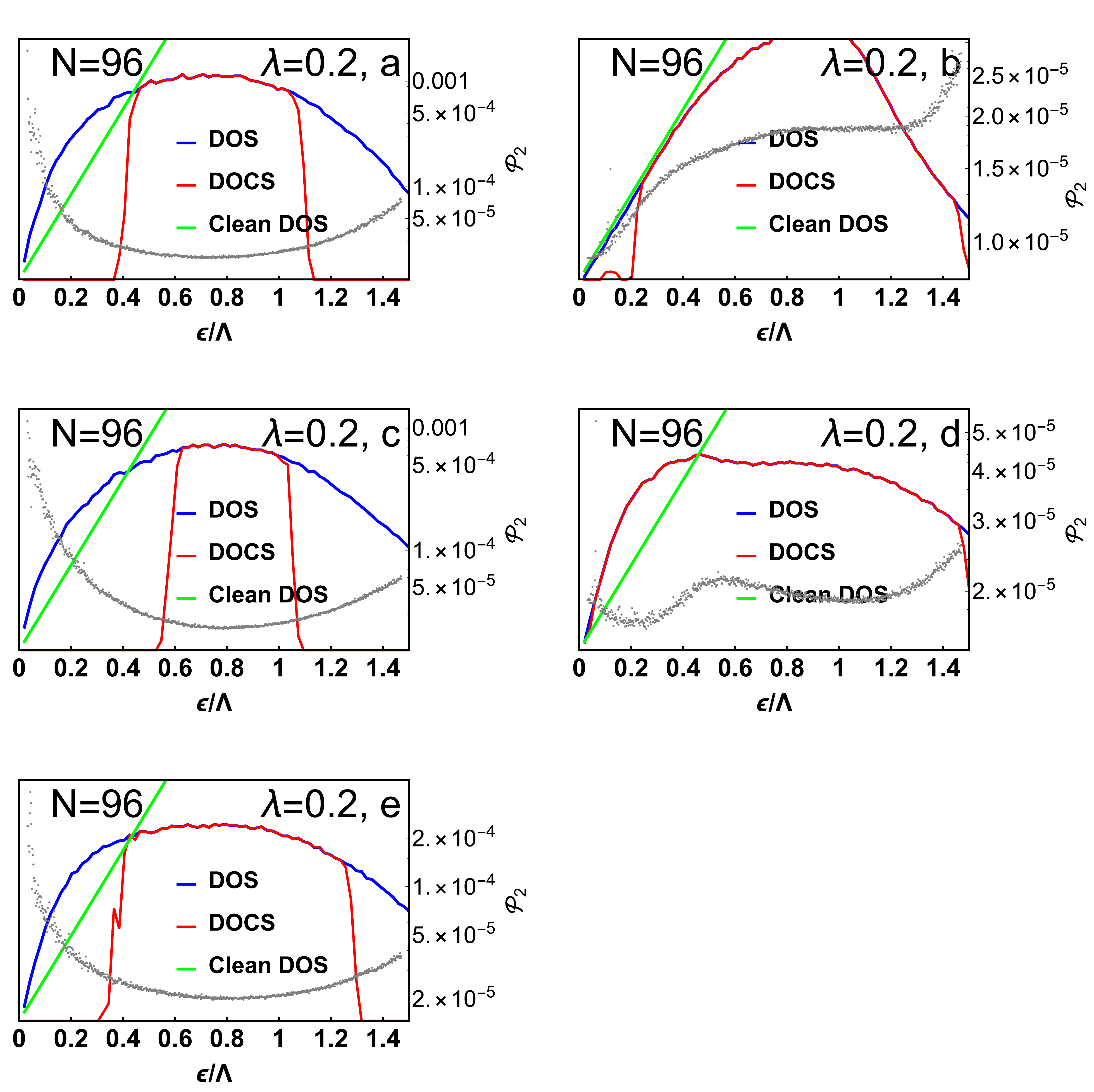}
	\caption{This figure shows the 
		total density of states (DOS)
		and the 
		density of critical states (DOCS)
		for the QGD, models (a)--(e), defined in and below Eq.~(\ref{DiracRVH}). 
		Results obtain by diagonalizing Hamiltonian from Eq.~(\ref{DiracRVH}) over a 
		$(2N+1)\times(2N+1)$ grid in momentum space, with $N = 96$ here \cite{Ghorashi2020}. 
		Data is plotted for the five different models at fixed dimensionless disorder strength $\lambda=0.2$;
		strong disorder corresponds to $\lambda \gtrsim 0.393$. 
		The DOCS counts the number of states with critical statistics (multifractal spectra) 
		that match a universal ansatz with a certain fitness criterion (see caption of Fig.~\ref{Fig--Dirt_Series_Dq(b)}).
		Also plotted is the second IPR $\mathcal{P}_2$ (gray dots), defined by Eq.~(\ref{eq:ipr}). 
		For models (a, c, e), a large swath of the spectrum appears critical for weak disorder. 
		However, as the disorder strength is increased, the swath shrinks \cite{Ghorashi2020}. 
		The IPR $\mathcal{P}_2$ shows that states outside of the swath are more rarified or localized
		than the critical ones. The linear-in-energy DOS of the clean limit is strongly distorted and filled-in at low energies, 
		which happens in models (a, c, e) for all $\lambda \gtrsim 0.2$. 
		These strong disorder effects are likely induced by the velocity component responsible 	
		for isotropic flattening or steepening of the cone.
		By contrast, models (b, d) show plane-wave states near zero energy for weak disorder. 
		Rarification near zero energy sets in for strong disorder; for model 	
		(d) the crossover is already visible at $\lambda=0.2$ shown here.
		The critical swath is larger and more robust to disorder strength in models (b, d), compared to (a, c, e). 
		Models (b, d) both contain nematic disorder and exclude isotropic flattening.
	}
	\label{Fig--DOS}
\end{figure}

\begin{figure}[b!]
	\centering
	\includegraphics[width=0.75\textwidth]{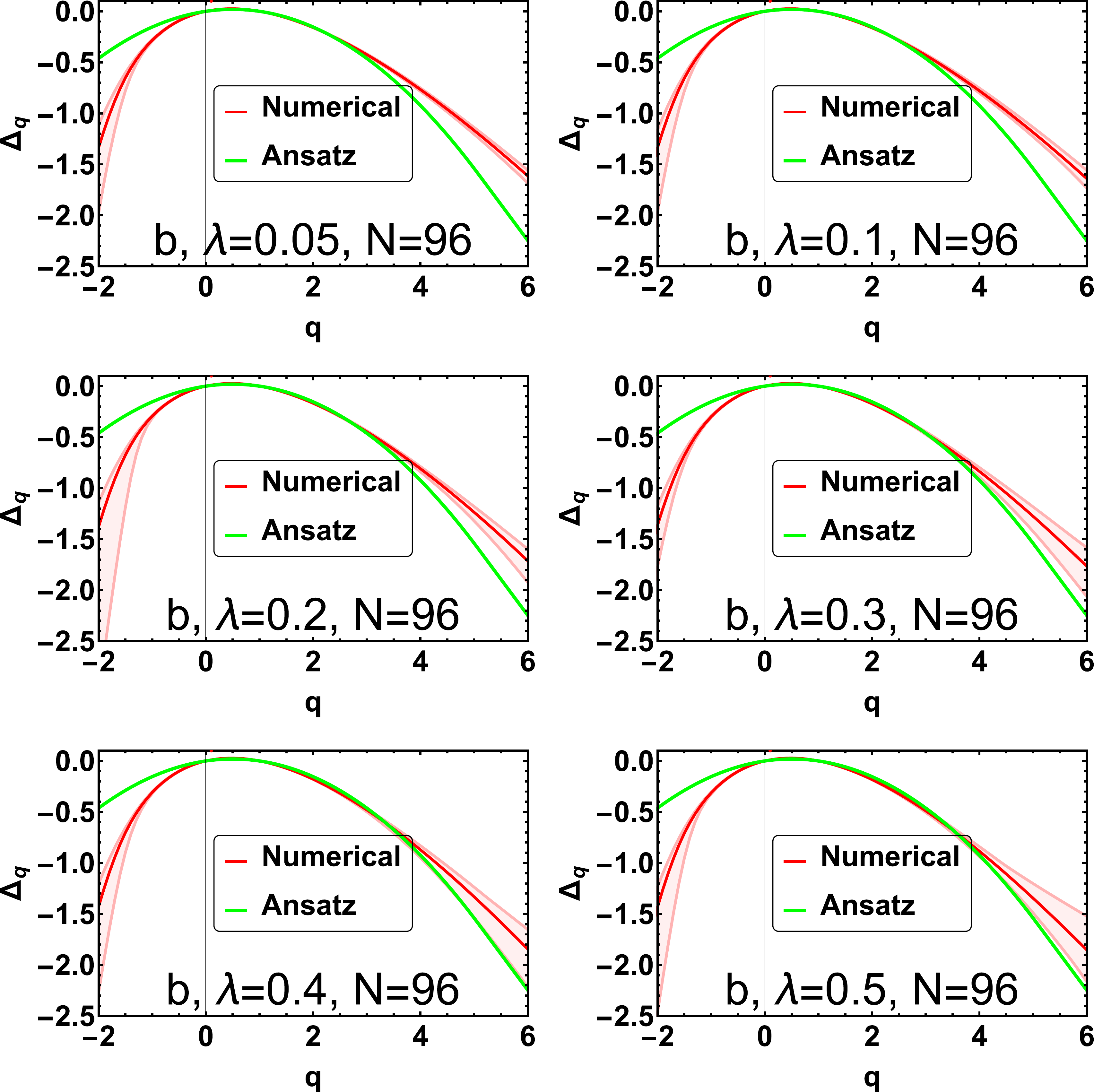}
	\caption{Anomalous multifractal spectrum $\Delta_q$ [Eq.~(\ref{Delta(q)Def})]
		for an energy bin of states selected from the DOS with the highest percentage of 
		critical states from Ref.~\cite{Ghorashi2020}. Here the spectrum is shown for model (b), 
		evaluated for the six different disorder strengths.
		The solid red curve denotes an average over the 15 states in the bin;
		the shaded red region indicates the standard deviation. 
		The green curve is the parabolic ansatz for $\Delta_q = -\theta \, q(1-q)$, 
		with $\theta = 1/13$. 
		States contributing to the critical count (DOCS) in Fig.~\ref{Fig--DOS}(b)
		match the parabolic ansatz within a certain threshold (see text) 
		over the range $0 < q \leq q_c = 5.1$.}
	\label{Fig--Dirt_Series_Dq(b)}
\end{figure}

Models (b) and (d) with \emph{nematic} QGD show the most robust
stacking of critical eigenstates with universal statistics, 
as reviewed below. In the context of the cuprates, this is interesting
because of the potential role of nematicity in these materials \cite{Fradkin2010,Agterberg2020}. 
In particular, evidence for \emph{quenched random} nematicity 
has emerged in recent studies of the pseudogap phase \cite{Davis2019}.

The natural generalization of the stacking conjectures in classes AIII and CI is
that the class DIII WZNW~$\rightarrow$~class D thermal quantum Hall plateau transition (TQHPT).
As reviewed in Sec.~\ref{sec:qh}, very little is known about the TQHPT. 
The multifractal spectra presented below may constitute the first (indirect) 
results substantiating a universal description of this transition.

\subsection{Multifractal analysis}

In Ref.~\cite{Ghorashi2020}, models (a)--(e) were studied for a large range of $N=32,\ldots,96$. 
In all five models, there are critical states at finite energy matching a $\theta = 1/13$ parabolic ansatz 
for $\Delta_q$ [Eqs.~(\ref{Delta(q)Def}) and (\ref{MFC_WZNW}), Table~\ref{tab:WZW}]. 
As for class CI in Sec.~\ref{sec:ci} and Fig.~\ref{fig:docs_ci}, 
we define the \emph{density of critical states} (DOCS) as follows. 
This is the proportion of states at finite energy matching 
the tolerance criterion that 85\% of the $\tau_q$ within the range $q\in[0,q_c]$ match the parabolic 
$\tau_q^\theta$ ansatz with $\theta = 1/13$ up to $4\%$ accuracy; here $q_c = 5.1$ is the termination
threshold \cite{Evers2008,Ghorashi2020,Chamon1996}.
The DOCS is plotted for all three models in Fig.~\ref{Fig--DOS}.

As the disorder strength is increased, the critical swath shrinks for model (a) and even more strongly for (c). 
In Fig.~\ref{fig:diii_metric} and in the definition below Eq.~\eqref{DiracRVH}, we can see that both of these 
models feature local isotropic flattening and steepening of the Dirac cone. 
Near zero energy in these models, there are stronger multifractal (rarified) states visible 
in the superimposed second IPR $\mathcal{P}_2$ (gray) in Fig.~\ref{Fig--DOS}. 
At high energies near the cutoff there is a crossover to localized Lifshitz tail states.

In the absence of the isotropic flattening and steepening, models (b), (d) instead exhibit 
an \emph{increasing} number of critical states (improved ``stacking'') with increasing disorder. 
Models (b) and (d) are similar, except that for the latter,  
rarified zero-energy states set in at intermediate disorder, see Fig.~\ref{Fig--DOS}. 
Indeed models (b) and (d) are related by local diffeomorphisms in the gravitational formulation of the problem \cite{Davis2020}. 
Since there are more independent disorder terms in (d), it is mapped to a (b) with effectively stronger disorder.
We note however that the interpretation of diffeomorphisms is different than in general
relativity or 2D quantum gravity \cite{KPZ,Distler,Z+Z}: 
at the surface of a dirty class DIII TSC, there is a preferred coordinate 
system $(x,y)$ that measures physical distances across the surface, given the flatness of 
physical spacetime. Disorder modulates the Majorana surface fluid in a way that is mathematically
identical to gravity, but geodesic distances are not directly measurable. 

Model (e) containing generic disorder exhibits a wide swath of critical states for weak disorder. 
However, this shrinks with increasing disorder strength, similar to models (a) and (c). 

The robustness of the anomalous multifractal spectra selected from an energy bin where the ratio of the DOCS to DOS is maximized
with respect to disorder is shown in Fig.~\ref{Fig--Dirt_Series_Dq(b)} for the exemplary model (b). Tuning over a whole order of 
magnitude in the disorder strength, the results are robust for $q$ not to close to 
the termination threshold $q_c$. The deviations are easily understood 
by the absence of ensemble averaging in the numerics.

Excitingly the behavior of models (b), (d) closely parallels observations from STM studies of BSCCO \cite{Davis08,DavisReview}. 
The cuprate superconductor shows quasiparticle interference at low energies, suggestive of plane wave states modified
by rare internode scattering, but strongly inhomogeneous, energy-independent spectra at higher energies. 
The inhomogeneous, energy-independent spectra could potentially be associated to the ``stacked'' multifractality,
robustly exhibited here for models with \emph{nematic} QGD \cite{Ghorashi2020}.


\section{Discussion}

The quantum-critical stacking conjecture for protecting class CI, AIII, and DIII
surface states from Anderson localization was outlined in detail in Sec.~\ref{sec:Fund}. 
The main predictions tested numerically are summarized in Table~\ref{tab:WZW}. 
Below we quickly recap the key results presented in Secs.~\ref{sec:aiii}--\ref{sec:diii}. 

It is worth emphasizing that at the surface of a 3D topological phase,
at energies of order the gap, the 2D surface states deconfine and 
merge with the bulk continuum. This is not captured by the pure continuum
2D Dirac theories studied here, which are instead implemented with a hard cutoff
in momentum space.
Boundary states of a bulk TSC lattice model in the slab geometry were 
studied in \cite{Sbierski2020}, with results consistent with the stacking
scenarios articulated in this review.

\paragraph{AIII}
The numerical analysis in Ref.~\cite{Sbierski2020} confirms the class AIII~$\rightarrow$~A IQHPT stacking prediction for $\nu=1$ \cite{Ostrovsky2007}. Multifractality in the surface theory, at the boundary of a bulk lattice model, and in the conductance distribution derived within the scattering matrix formalism where shown to match the universal IQHPT characteristics. 

Notably, however, the even-odd effect implied by the NL$\sigma$M-based derivation in Ref.~\cite{Ostrovsky2007} is absent [see also Eqs.~(\ref{WZNW}) and (\ref{Pruisken}), above]. Remarkably, both $\nu=1,2$ show a stack of IQHPT critical states at finite energy. 

Our computations of the Kubo $\sk^{xx}$ conductivity add support to these results.

\paragraph{CI}
In Ref.~\cite{Ghorashi2018}, time-reversal invariant TSCs with full spin-rotation symmetry in class CI were investigated for both small and large winding numbers $\nu$. From the multifractal spectra, the stacking conjecture was posited for finite-energy surface states, 
CI WZNW~$\rightarrow$~class C spin QHPT. Here we reproduced some of the results for the minimal winding number $\nu=2$ in larger systems to show that these observations are robust with respect to finite-size effects.

We were able to additionally show that the finite-energy Kubo conductivity is reasonably close to the class-C SQHPT prediction (see Table~\ref{tab:WZW}). Most important, it tends towards a finite value, in contrast to the conventional expectation of Anderson localization.

\paragraph{DIII}
Finally, the generic time-reversal topological superconductor with no spin symmetry at all residing in class DIII also shows clear signs of stacking at finite energy. In Ref.~\cite{Ghorashi2020}, reviewed in Sec.~\ref{sec:diii}, only the minimal $\nu = 1$ single Majorana surface
cone is considered. In this case, time-reversal invariant disorder couples to the stress tensor, producing ``quenched gravitational disorder'' (QGD). This type of disorder should be present in any 2D massless Dirac material, but is strongly irrelevant at the Dirac point and typically overwhelmed by other types of disorder. 

In claiming stacked criticality for class DIII surface states, extra care is required. The conventional picture articulated in 
Secs.~\ref{sec:topclass} and \ref{sec:wzw} is that finite-energy states for a class DIII Hamiltonian belong in the Wigner--Dyson symplectic class AII. In 2D, this class shows weak antilocalization with a logarithmic scaling in the system size. Logarithms are notoriously hard to observe in finite-size studies; however, the multifractal spectra should show a strong dependence on 
the disorder strength 
$\lambda$ (since it determines the bare value of the conductance in the conventional scenario). 
For QGD, the finite-energy multifractality is found to be approximately parabolic, and the spectra change little despite the variation of the disorder strength $\lambda$ by an order of magnitude. 

In analogy with classes CI and AIII, the natural stacking conjecture in this case relates DIII WZNW~$\rightarrow$~the class D thermal quantum  Hall plateau transition (TQHPT).
Although it is natural to expect a TQHPT with universal multifractal and conductance statistics, no such observation or calculation has yet succeeded indicating this kind of transition in class D, see Sec.~\ref{sec:qh}.

The coexistence of low-energy plane-waves and finite-energy critical states (with fixed, universal multifractal fluctuations in the latter) 
over a wide range of disorder strengths means that model (b) is similar to the phenomenology of STM observations in the high-$T_c$ cuprate 
BSCCO \cite{Ghorashi2020,Davis01,Davis02,Davis05-a,Davis05-b,Davis08,DavisReview}.
Although not a topological system, the low-energy Dirac quasiparticle description of a 2D $d$-wave superconductor reduces to independent ``topological'' components when interpair and/or internode scattering is suppressed by hand \cite{Ghorashi2020,AltlandSimonsZirnbauer2002}.
The critical stacking for a single cone is found to be most robust for nematic QGD. 

A remarkable property of the class CI, AIII, and DIII WZNW topological surface theories is the precisely quantized longitudinal surface spin or thermal conductivity at zero energy, which is robust to the presence of \emph{both} disorder and interactions \cite{Evers2008,Tsvelik1995,Ludwig1994,Xie2015,Ostrovsky2006}.
We reviewed studies \cite{Ghorashi2020,Sbierski2020,Ghorashi2018} augmented by unpublished numerical evidence indicating an even more surprising connection between 2D classes C, A, and D quantum Hall effects and 3D classes CI, AIII, and DIII topological superconductors.


\section{Outlook and further directions}

There are several numerical open tasks: 
\begin{enumerate}
\item{
Computing the finite-energy Landauer conductance in the other classes CI, DIII. 
In class CI one can compare against the Cardy result \cite{SQHPT-4_Cardy2000} for the class C network model conductance.
}
\item{
Analyzing higher winding numbers in class DIII systems. The winding $\nu=1$ case is in some aspects very different from 
the WZNW models at higher $\nu$. An important question is whether the observed $\theta_{\e\neq 0}=0.077$ (Table~\ref{tab:WZW})
is also robust to changes in the winding number.}
\item{
Increase system sizes using Arnoldi techniques instead of full diagonalization. The network models for QH effects can be studied 
at more than an order of magnitude larger linear sizes $L = \mathcal{O}(2000)$. Network models can be described by sparse matrices. 
What actually enters the algorithm is the (usually not sparse) resolvent operator of the Hamiltonian, that can be computed efficiently for sparse Hamiltonians. For generic dense matrices computing the resolvent is as hard as full diagonalization. 
Using the banded structure of the Hamiltonian matrix in $k$-space, it is possible to construct the resolvent operator faster 
than in the generic case.} 
\item{
We need to find evidence for the existence of the conjectured thermal quantum Hall plateau transition in class D in the presence of disorder, and compute the thermal conductivity and multifractal spectrum in order to 
compare with the stacked criticality in our DIII results. 
}
\end{enumerate}

The current analytical understanding of the stacking phenomena observed numerically is 
primitive. Goals include the following:
\begin{enumerate}
\item{Ref.~\cite{Essin2011} derives an interesting form of bulk-boundary correspondence for all symmetry classes. A winding number of the critical surface states is constructed and shown to be equal to the winding number of the bulk. A fruitful direction of research would be to investigate the applicability of these concepts to the critical stacks at finite energies.} 
\item{Another goal is to improve the NL$\sigma$M expansion that gives the class AIII WZNW~$\rightarrow$~A IQHPT prediction for odd winding numbers.} 
\item{Gain better understanding of the class  DIII gravitational theory \cite{Davis2020}.}
\item{Investigate the effects of interparticle interactions on stacked critical states.}
\end{enumerate}

Finally, a key problem is to understand the depth of the relationship implied by the stacking of classes A, C, D topological quantum phase transitions (the Hall plateaux transitions) at the surface of classes AIII, CI, and DIII TSCs. Although this connection has been revealed indirectly here, through studies of the effects of disorder on TSC surface theories, it suggests a more intrinsic, topological relationship between these classes (which govern topological phases with integer-valued invariants in two and three dimensions). 
One idea is the following. Is it possible to reproduce the critical statistics studied here by studying an entanglement cut \cite{Haldane2008} 
for a \emph{clean} bulk TSC Hamiltonian? Instead of averaging over disorder configurations at a physical surface, perhaps
one can average over \emph{bulk quasiparticle band structures}.

\section{Acknowledgments}

We thank our coauthors on the work reviewed here \cite{Ghorashi2020,Sbierski2020,Ghorashi2018}: 
S.\ A.\ A.\ Ghorashi, 
Y.\ Liao,
B.\ Sbierski, 
and 
S.\ M.\ Davis. 
JFK acknowledges funding by Graduate Funding from the German States awarded by KHYS.
MSF acknowledges support by NSF CAREER Grant No.~DMR-1552327.

\end{document}